\documentclass[prd,superscriptaddress,amsfonts,amssymb,amsmath,showpacs,twocolumn]{revtex4-2}
\usepackage{bm}
\usepackage{amsfonts}
\usepackage{latexsym}
\usepackage[latin1]{inputenc}
\usepackage{graphicx}
\usepackage{amsmath}
\usepackage{palatino}
\usepackage{mathpazo}
\usepackage{textcomp}
\usepackage{float}
\usepackage{booktabs}
\usepackage{dcolumn}
\usepackage{ragged2e}
\usepackage{hyperref}
\hypersetup{colorlinks,citecolor=red}
\hypersetup{colorlinks=true,linkcolor=blue,filecolor=blue,    urlcolor=magenta}
\usepackage{amsmath}
\usepackage{xcolor}
\usepackage[caption=false]{subfig}
\usepackage{commath}
\usepackage{csquotes}

\def\jnl@style{\it}
\def\aaref@jnl#1{{\jnl@style#1}}

\def\aaref@jnl#1{{\jnl@style#1}}

\def\aj{\aaref@jnl{AJ}}                   
\def\apj{\aaref@jnl{ApJ}}                 
\def\apjl{\aaref@jnl{ApJ}}                
\def\apjs{\aaref@jnl{ApJS}}               
\def\apss{\aaref@jnl{Ap\&SS}}             
\def\aap{\aaref@jnl{A\&A}}                
\def\aapr{\aaref@jnl{A\&A~Rev.}}          
\def\aaps{\aaref@jnl{A\&AS}}              
\def\mnras{\aaref@jnl{Mon.~Not.~Roy.~Astron.~Soc.}}             
\def\prd{\aaref@jnl{Phys.~Rev.~D}}        
\def\plb{\aaref@jnl{Phys.~Lett.~B}}        
\def\prc{\aaref@jnl{Phys.~Rev.~C}}  
\def\prl{\aaref@jnl{Phys.~Rev.~Lett.}}    
\def\qjras{\aaref@jnl{QJRAS}}             
\def\skytel{\aaref@jnl{S\&T}}             
\def\ssr{\aaref@jnl{Space~Sci.~Rev.}}     
\def\zap{\aaref@jnl{ZAp}}                 
\def\nat{\aaref@jnl{Nature}}              
\def\aplett{\aaref@jnl{Astrophys.~Lett.}} 
\def\apspr{\aaref@jnl{Astrophys.~Space~Phys.~Res.}} 
\def\physrep{\aaref@jnl{Phys.~Rep.}}      
\def\physscr{\aaref@jnl{Phys.~Scr}}       
\def\commat{\aaref@jnl{Comm.~Math.~Phys.}}              
\def\science{\aaref@jnl{Science}}               
\def\cqg{\aaref@jnl{Classical Quant.~Grav.}}            
\def\jpcs{\aaref@jnl{JPCS}}                                     
\def\ijmpd{\aaref@jnl{Int.~J.~Mod.~Phys.~D}}                    
\def\grg{\aaref@jnl{Gen.~Relat.~Gravit.}}               
\def\rpp{\aaref@jnl{Rep.~Prog.~Phys.}}          
\def\npa{\aaref@jnl{Nucl.~Phys.~A}}        
\def\lrr{\aaref@jnl{Living Rev.~Rel.}}                   
\def\jcap{\aaref@jnl{J.~Cosmology Astropart.~Phys.}}    
\def\rmp{\aaref@jnl{Rev.~Mod.~Phys.}}   
\def\epjc{\aaref@jnl{Eur.~Phys.~J.~C}}

\allowdisplaybreaks[1]

\begin{document}

\color{black}       

\title{Dark matter admixed relativistic stars: Structural properties and tidal Love numbers}

\author{Takol Tangphati} 
\email[]{takoltang@gmail.com}
\affiliation{School of Science, Walailak University, Thasala, \\Nakhon Si Thammarat, 80160, Thailand}
\affiliation{Research Center for Theoretical Simulation and Applied Research in Bioscience and Sensing, Walailak University, Thasala, Nakhon Si Thammarat 80160, Thailand}

\author{Grigoris Panotopoulos}
\email{grigorios.panotopoulos@ufrontera.cl}
\affiliation{Departamento de Ciencias F{\'i}sicas, Universidad de la Frontera, Casilla 54-D, 4811186 Temuco, Chile}

\author{Ayan Banerjee} \email[]{ayanbanerjeemath@gmail.com}
\affiliation{Atrophysics Research Centre, School of Mathematics, Statistics and Computer Science, University of KwaZulu--Natal, Private Bag X54001, Durban 4000, South Africa}

\author{Anirudh Pradhan} 
\email[]{pradhan.anirudh@gmail.com}
\affiliation{Centre for Cosmology, Astrophysics and Space Science, GLA University, Mathura-281 406, Uttar Pradesh, India}


\date{\today}

\begin{abstract}
We study the impact of bosonic, self-interacting dark matter on structural properties and tidal deformabilities of compact stars. As far as the gravitational theory is concerned, we assume Einstein's gravity in four dimensions with a vanishing cosmological constant. Regarding matter content, we consider a state-of-matter to a linear form of equation-of-state (EoS), while for dark matter we assume a quartic scalar potential, which implies a certain non-linear EoS obtained long time ago. Adopting the two-fluid formalism we integrate the structure equations as well as the Riccati equation for the metric even perturbations imposing appropriate initial conditions at the center of the stars and matching conditions at their surface. We compute the stellar mass and radius, factor of compactness and dimensionless deformability varying several free parameters of the model studied here. Tidal deformability and the corresponding tidal Love number determine the imprint of the underlying EoS within the signals emitted during binary coalescences, and it is expected to be altered due to the presence of dark matter inside the objects. We find that in all cases considered here, the dimensionless deformability of the canonical stellar mass remains lower than the upper bound, $\Lambda_{1.4} < 800$. We also look at the stability of these stars based on the Harrison-Zeldovich-Novikov criterion under various conditions. It is observed that the presence of dark matter implies significantly lower highest stellar mass, and also smaller and more compact stars for a given stellar mass.
\end{abstract}

\maketitle

\section{Introduction}

Observable signals from the merging of compact stars, such as those involving black holes and neutron stars, significantly enhance our understanding of their internal characteristics and evolutionary dynamics. This development consequently opens up a new way for astronomy to engage a broader audience. Moreover, the detection of gravitational waves from binary neutron star (BNS) mergers \cite{LIGOScientific:2017vwq}, along with precise radius measurements from Neutron Star Interior Composition Explorer (NICER) \cite{Miller:2021qha}, 
has offered unparalleled insights into the neutron star (NS) equation of state (EoS), whether they are composed of conventional nuclear matter or contain more exotic forms of matter. In this line, a comprehensive study regarding the composition of stellar matter, incorporating various hypotheses that may predict differing structural properties of compact stars (CSs), has been found in Refs. \cite{Lattimer:2004pg,Alford:2004pf,Naz:2024aot,Gholami:2024ety,Malik:2024yex,Odintsov:2023ypt,Naz:2025fzh}.

Among various possibilities, dark matter (DM)-whether it exists as a bosonic or fermionic particle-may combine with ordinary matter to form a new type of compact object known as DM-admixed CSs. The origin and nature of DM remain among the most enthralling topics in astrophysics and cosmology over the last few decades
 \cite{Ciarcelluti:2010ji,Henriques:1990xg,Raj:2017wrv}. As we know, CSs in galaxies are viewed as natural laboratories
where theories can be tested and observational data can be collected. Thus, we speculate that several adjacent stellar entities, specifically black holes, proto-neutron stars, supernovae and their remnants, may play a crucial role  in the search for
dark matter \cite{Bramante:2023djs}. Furthermore, the presence of DM within compact objects may provide significant insights into its behavior and properties,
thereby advancing the understanding of this enigmatic phenomenon. Given its richness, various models of DM have suggested its presence in those stars including the study of accumulation within stars, its changes their internal structure and causes a shift of the tidal deformability, which could be observable in the gravitational-wave (GW) signals of BNS mergers \cite{Bertone:2007ae,Kouvaris:2010vv,Brito:2015yga,Bell:2013xk,Liu:2023ecz,Lopes:2023uxi,Panotopoulos:2018ipq,Routaray:2022utr,Panotopoulos:2017idn,Malik:2025raq}.  This suggests that NSs serving as laboratories for indirectly measuring the properties of DM, thereby offering a new perspective on the study of the
evaluation history of compact astrophysical objects.

Another important question related to this topic is how DM could mix with ordinary matter inside a NS, as described in Refs. \cite{Zheng:2014fya,Kouvaris:2007ay}. Furthermore, previous studies have explored various possibilities through which celestial bodies, such as NSs or QSs, can accumulate DM and interact with nuclear/quark matter, as discussed in Refs. \cite{Gani:2018mey,Nguyen:2022zwb,Voskresensky:2022fzk}. Therefore, it is useful to consider a two-fluid system in which DM, either as a bosonic or fermionic particle, can be mixed with ordinary matter, which leads to significant effects on its mass-radius relationship and dimensionless tidal deformability ($\Lambda$) \cite{Lourenco:2022fmf,Giangrandi:2022wht,Naz:2024uip,Thakur:2023aqm} to understand the microscopic properties of the matter that forms them.   

 Extracting information on the inner structure and composition of relativistic stars is one of the primary goals of current and future gravitational wave detectors. The inspiral and relativistic collision of two compact objects in a binary system, and the gravitational wave signal emitted during the process, contain a wealth of information on the nature of the colliding bodies. Furthermore, in a binary system one of the stars is subjected to the external gravitational field produced by the companion object. The imprint of the EoS within the signals emitted during binary coalescences is mostly determined by adiabatic tidal interactions, characterized in terms of a set of coefficients,
known as the tidal deformability and the corresponding tidal Love numbers. The theory of tidal deformability was first introduced in Newtonian gravity over one century ago
by Love \cite{Love1,Love2}, with the purpose of understanding the yielding of the Earth to disturbing forces. For
a spherical body, Love introduced two dimensionless numbers to describe the tidal response of the
Earth. The first one, $h$, describes the relative deformation of the body in the longitudinal direction (with
respect to the perturbation); the other one, $k$, describes the relative deformation of the gravitational potential.


The goal of this work is to study the two-fluid formalism in GR to numerically solve the Tolman-Oppenheimer-Volkoff (TOV) equations. Particularly,
we investigated the macroscopic properties of CSs, including their maximum masses and radii, as well as the tidal deformability of 
CSs composed of a combination of dark matter and ordinary matter, referred to as DM admixed compact stars. In this formalism, we explored the impact of self-interacting bosonic dark matter (DM) on the structure and properties of CSs under different model parameters. We further assumed that the DM primarily interacts with the barotropic equation of state (EoS) 
and calculated the mass, radius, and tidal deformability for the DM-admixed CS. Obtained  results are then constrained using the observational data from the GW170817 event.  Additionally, we conducted a stability analysis of these two-fluid objects by examining the static stability criterion, the adiabatic index, and the sound speed.  The properties of NSs/QSs admixed with self-interacting bosonic dark matter have been  examined in Ref. \cite{Pitz:2024xvh,Buras-Stubbs:2024don,Dutta:2024vzw}.

 Through the application of these concepts, it would be our desire and aim to extend the frontiers of knowledge regarding compact stellar objects and their essential physical attributes. To our knowledge, this study analyzes the effects of dark matter on compact stellar objects using a two-fluid approach.  In this investigation, the obtained results are expressed in natural units (  $\hbar = c = G = 1$ ), and the structure of our presentation is as follows: In Sec.~\ref{sec2}, we describe the two classes of EoSs
 to describe the dark sector and ordinary matter in our analysis.  In Sec.~\ref{sec3}, we present the structural equations that cover the theoretical framework used in this study. The results are discussed in Sec.~\ref{sec4}, where we present the mass-radius and mass-compactness relations for DM admixed compact stars. In Sec.~\ref{sec5} we study the tidal deformability of CSs and show that in this case, DM admixed CSs are ruled out by the astrophysical observations. In Sec.~\ref{sec6} we examine the stability of the configuration. Finally, we summarize our paper in Sec.~\ref{sec7} and propose future research directions.

\section{Two-fluid Formalism}\label{sec2}

Considering the growing interest in the interior properties of compact objects, it is essential to define the conditions that pertain to the matter located in their interiors. Thus, in the next two subsections, we will provide an overview that describes first barotropic EoS and then normal matter admixed with self-interacting bosonic dark matter.

\subsection{Linear equation of state}

Based on physical grounds, we expect that the distribution of matter for realistic stellar objects should satisfy a barotropic EoS $p=p(\rho)$. For our purposes, we presume the linear EoS \cite{Sharma:2007hc,Takisa:2014sva,Hussain:2025gvx}
\begin{eqnarray}\label{eq1}
   p_q = \alpha \left(\rho_q-\rho_s\right),
\end{eqnarray}
where $0< \alpha< 1$ represents a constant that is associated with the sound speed $dp/d\rho = \alpha$, and $\rho_s$ denotes the density at the stellar surface defined by $r = s$, at which the pressure vanishes.

\subsection{Equation of state for dark matter}

The accretion of DM influences the characteristics of compact objects in two significant ways. Firstly, the process of DM annihilation
may increase the temperature of the compact object, thereby affecting its kinematic properties,  including both linear and angular momentum \cite{Perez-Garcia:2011tqq,deLavallaz:2010wp}. Here, we adopt the DM particles as massive self-interacting bosons, and we employ a commonly referenced generic DM model characterized by a single parameter, $\rho_0$, to represent the EOS for DM, which can be formulated as \cite{Karkevandi:2021ygv,Colpi:1986ye}: 
\begin{equation}\label{eq2}
p_d=\frac{4}{9} \rho_{0} \left(\sqrt{1+\frac{3\rho_d}{4\rho_{0}}}-1\right)^{2}
\end{equation}
in the strong-coupling limit \cite{Colpi:1986ye}, where $\rho_{0}= m_{\chi, b}^{4} / 4\lambda $, and $m_{\chi, b}$, $\rho$, and $\lambda$ are the particle mass, density, and coupling constant of the self-interacting bosonic DM. Since the nature of DM is still unknown to us, there is no limit to the possible values of $m_{\chi,b}$ \cite{Kouvaris:2010jy,Bramante:2013hn}. In the following, we consider four typical masses of $m_{\chi,b}$ = 100, 150, 200 and 250 MeV \cite{Lopes:2018oao}.  Furthermore, we choose $\lambda = \pi$ as shown in Ref.~\cite{Karkevandi:2021ygv}, as its exact value is less importance due to $\lambda$ entering with fourth power in the definition of $\rho_0$.  It's important to note that
 the positive $\lambda$ implies the repulsive self-interaction between DM particles, which stabilizes pure boson dark stars against strong gravity.


\section{Structural Equations}\label{sec3}

Given the interest in studying compact objects resulting from the admixture of ordinary matter and self-interacting bosonic dark matter, we write down a two-fluid version of the TOV equations \cite{Sandin:2008db,Comer:1999rs}:
\begin{eqnarray}
 \frac{dp_d}{dr} &= -[p_d + \rho_d]\frac{d\nu}{dr} \:,
\\
 \frac{dp_q}{dr} &= -[p_q + \rho_q]\frac{d\nu}{dr} \:,
\label{e:tovn}
\\
 \frac{dm}{dr}   &= 4\pi r^2 \rho \:,
\label{e:tovm}
\\
 \frac{d\nu}{dr} &= \frac{m + 4\pi r^3p}{r(r - 2m)} \:,
\label{e:tovnu}
\end{eqnarray}
with $r$ being the radial coordinate from the center of the star, and $p=p_q+p_d$, $\rho=\rho_q+\rho_d$, $m=m_q+m_d$
are the total pressure, energy density, and mass of the two components at radius $r$, respectively. We then have for the total mass of the star,
\begin{eqnarray}
 M = m_q(R_q) + m_d(R_d) \:.
 \label{eq:gravmass}
\end{eqnarray}

Additionally, the total DM mass as a fraction of the total mass of the DM admixed CSs is then
\begin{eqnarray}
 f = \frac{m_d(R_d)}{M} \:,
 \label{eq:gravmass}
\end{eqnarray}
where the stellar radii $R_q$ and $R_d$
are defined by the vanishing of the respective pressures.


\smallskip

\section{Numerical Results and Discussions}\label{sec4}

In the following, we delve into the existence of CSs admixed with bosonic dark matter on the mass-radius relations and determine the parameter space of DM particle mass and fraction as well. Utilizing the EoSs provided in (\ref{eq1}) and (\ref{eq2}), we search for the physical solution by solving the set of ODEs (\ref{e:tovm})-(\ref{e:tovnu}). The numerical outputs are based on initial conditions set at the star's core $(r = 0)$, where $m(0) = 0$ and $\rho(0) = \rho_c$, with $\rho_c$ denoting the central energy density. Moreover, we define the boundary conditions $p(R) = 0$ at which the fluid pressure vanishes at the star's surface $(r = R)$. To begin with, we select four sets of parameters $(\alpha$,  $m_{\chi,b}$, $F_x$ ~and~ $\rho_s)$ to see their impact on the stellar properties, thereby covering a wide range of cases that may correspond to CSs. 

\begin{figure}[h]
    \centering
    \includegraphics[width = 8 cm]{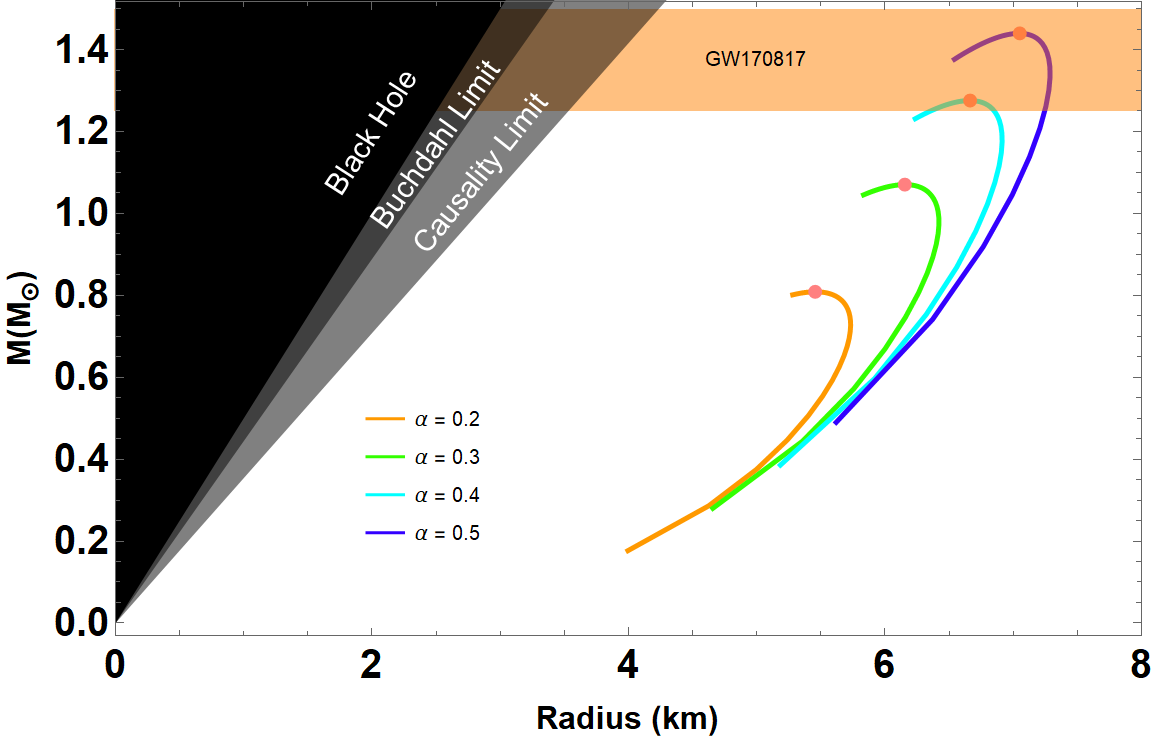}
    \includegraphics[width = 8 cm]{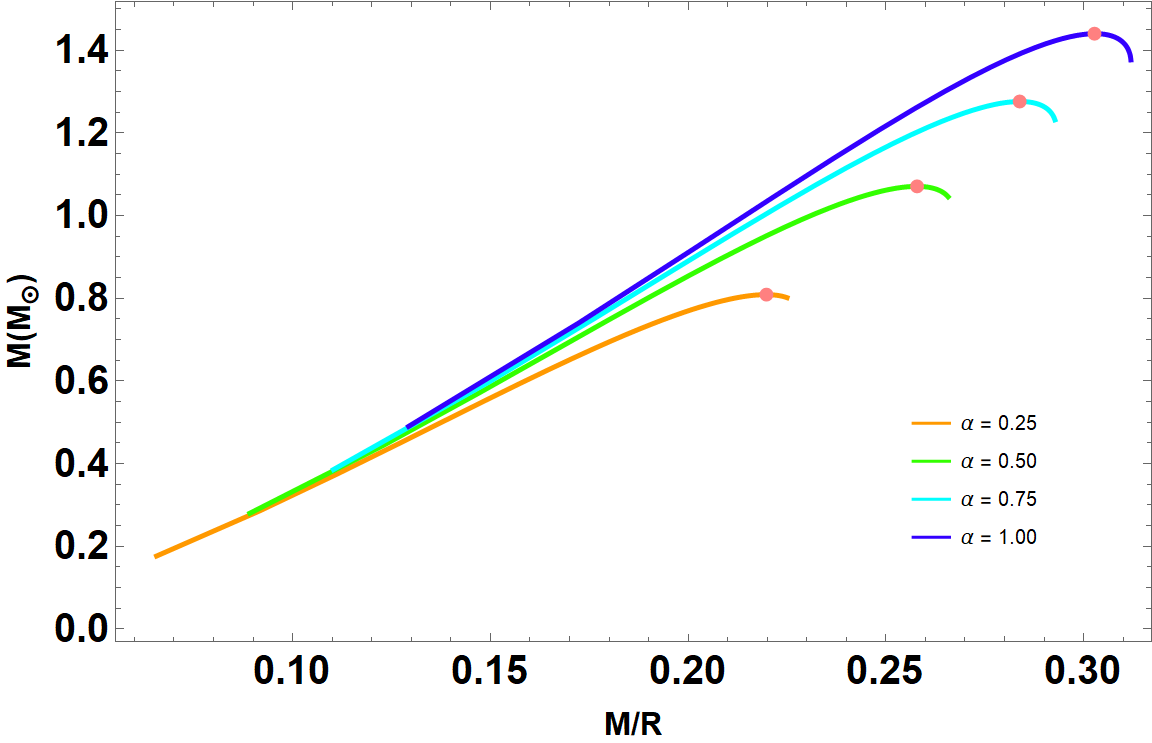}
    \caption{The mass-radius $(M-R)$ and mass-compactness $(M-M/R)$ profiles for DM-admixed compact stars with different values of the dimensionless constant $\alpha \in [0.2, 0.5]$. The other parameter sets are $m_{\chi,b} = 200$ MeV, $F_X =\frac{\rho_N}{\rho_D} = 0.95$ and $\rho_s = 1.17 \times 10^{15}$ g/cm$^{3}$, respectively. The mass-radius relation is shown in conjunction with the constraints from the GW170817 merger event \cite{LIGOScientific:2018cki}. }
    \label{fig_vary_alpha}
\end{figure}

\begin{figure}[h]
    \centering
    \includegraphics[width = 8 cm]{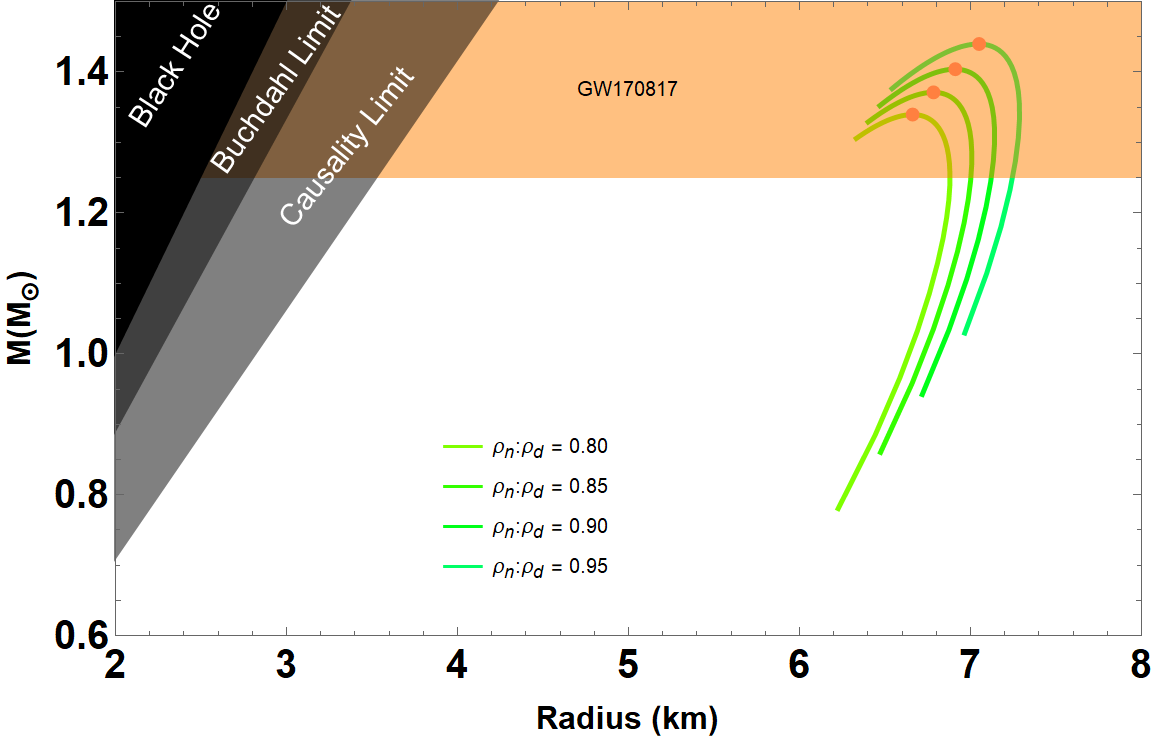}
    \includegraphics[width = 8 cm]{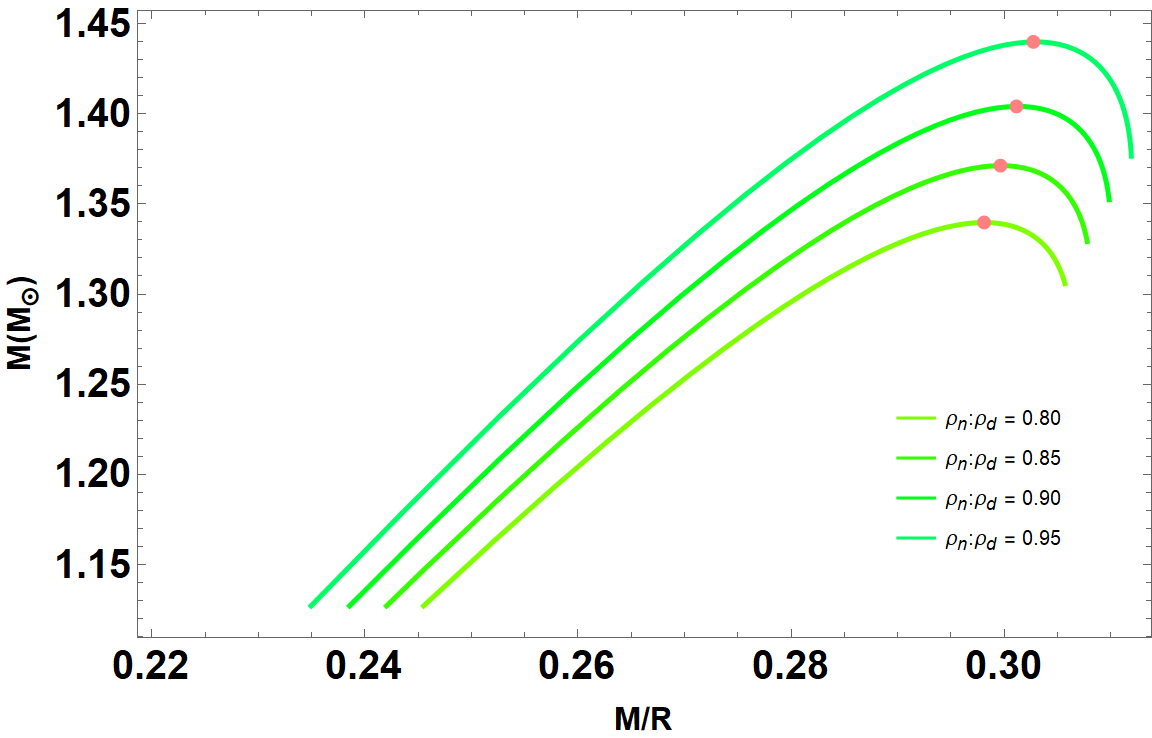}
    \caption{The  $(M-R)$ and $(M-M/R)$ profiles for DM-admixed compact stars with different values of the dimensionless constant $F_x \in [0.8, 0.95]$. The other parameter sets are $m_{\chi,b} = 200$ MeV, $\alpha = 0.5$, and $\rho_s = 1.17 \times 10^{15}$ g/cm$^{3}$, respectively.}
    \label{fig_vary_Ratio}
\end{figure}

\begin{figure}[h]
    \centering
    \includegraphics[width = 8 cm]{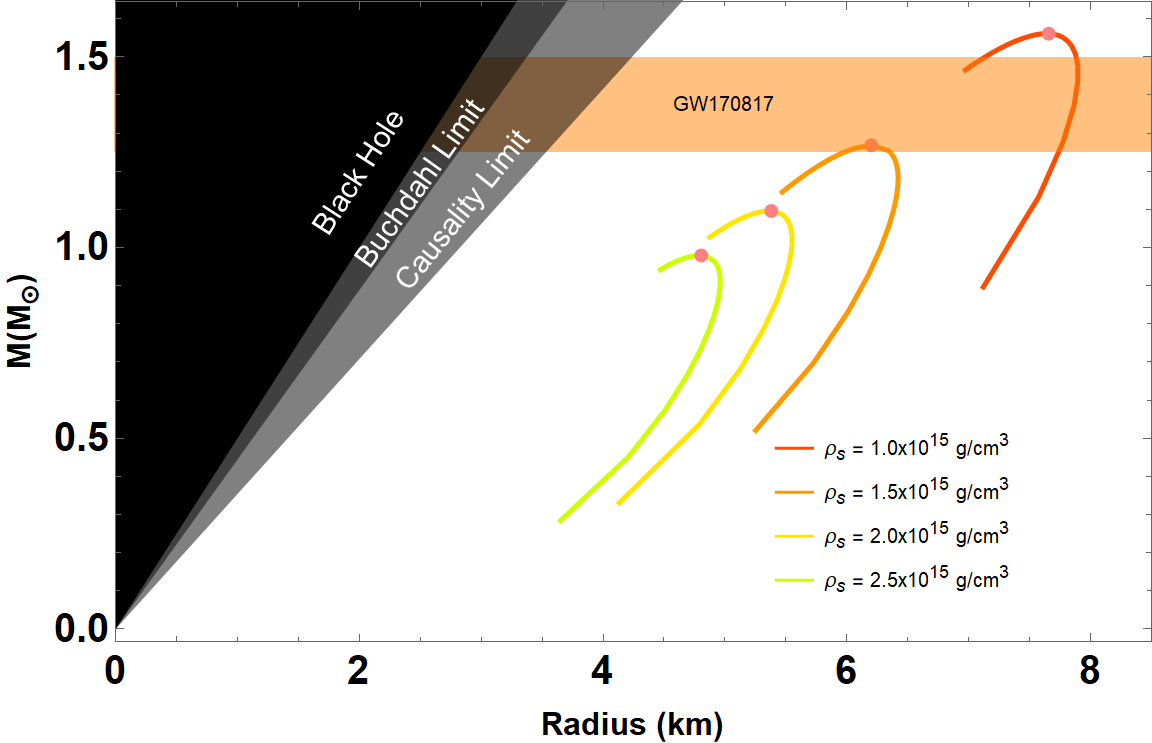}
    \includegraphics[width = 8 cm]{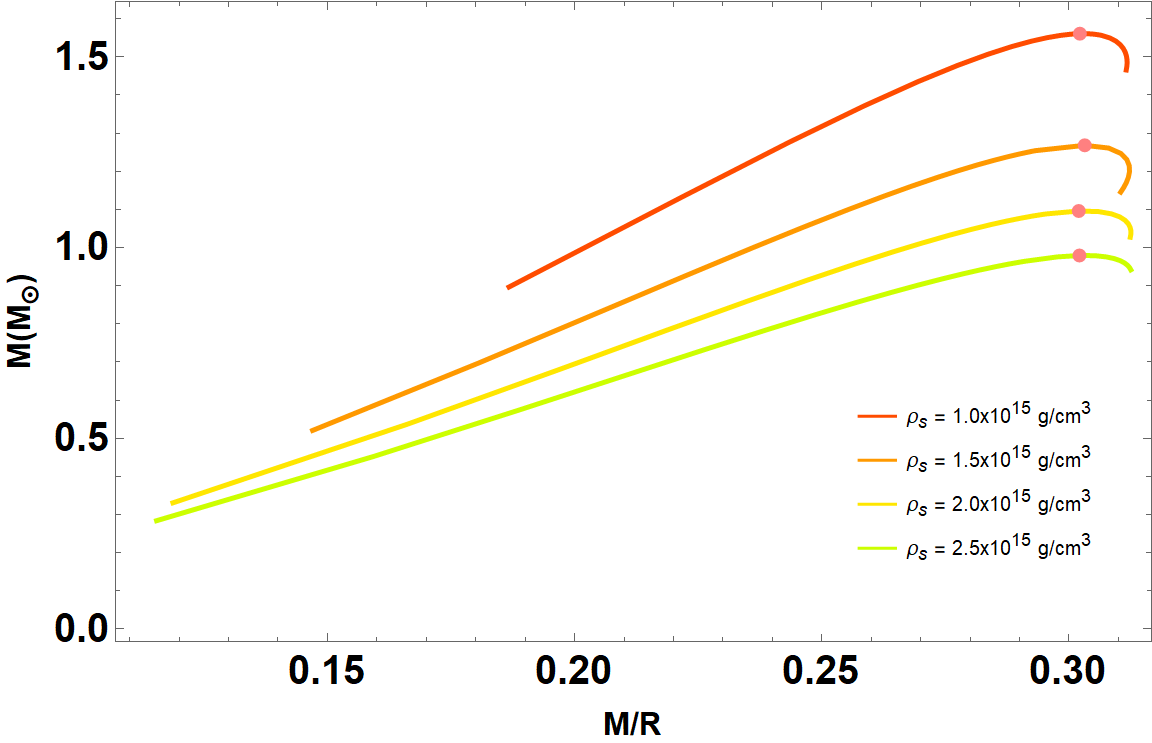}
    \caption{The $(M-R)$ and $(M-M/R)$ profiles for DM-admixed compact stars with different values of the dimensionless constant $\rho_s  \in [1.0, 2.5] \times 10^{15}$ g/cm$^{3}$.  The other parameter sets are $m_{\chi,b} = 200$ MeV, $\alpha = 0.5$, and $F_x = 0.95$, respectively.}
    \label{fig_vary_rhos}
\end{figure}

\begin{figure}[h]
    \centering
    \includegraphics[width = 8 cm]{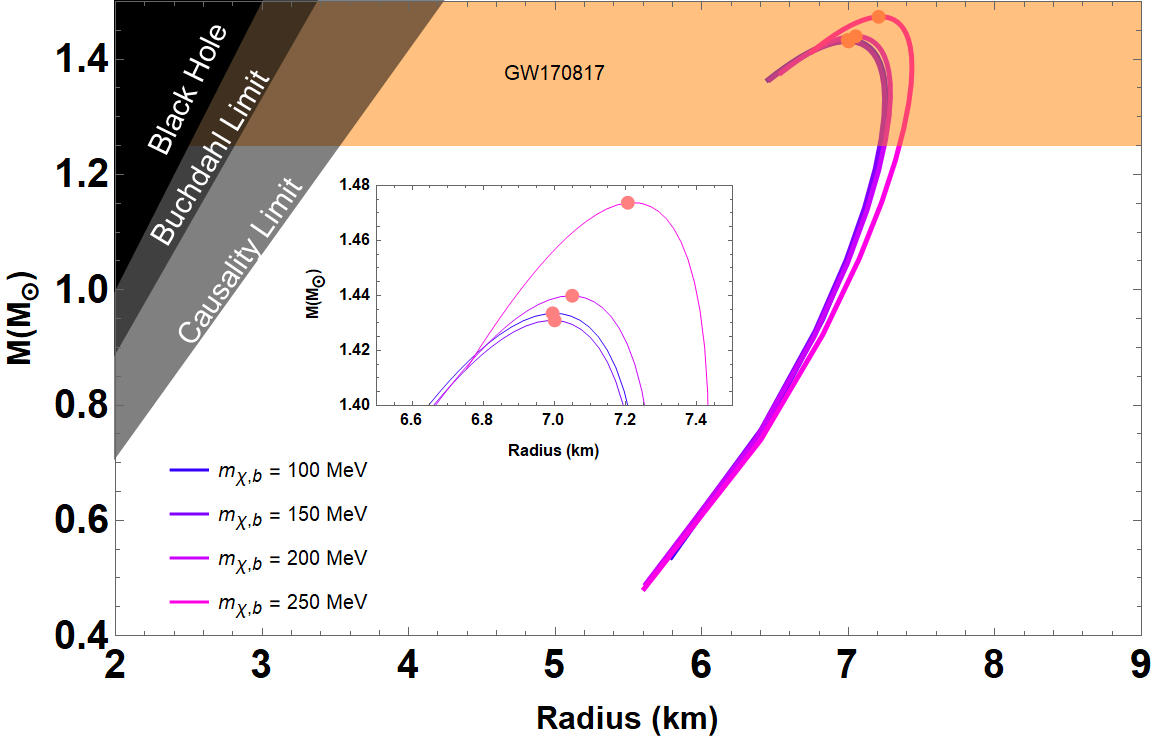}
    \includegraphics[width = 8 cm]{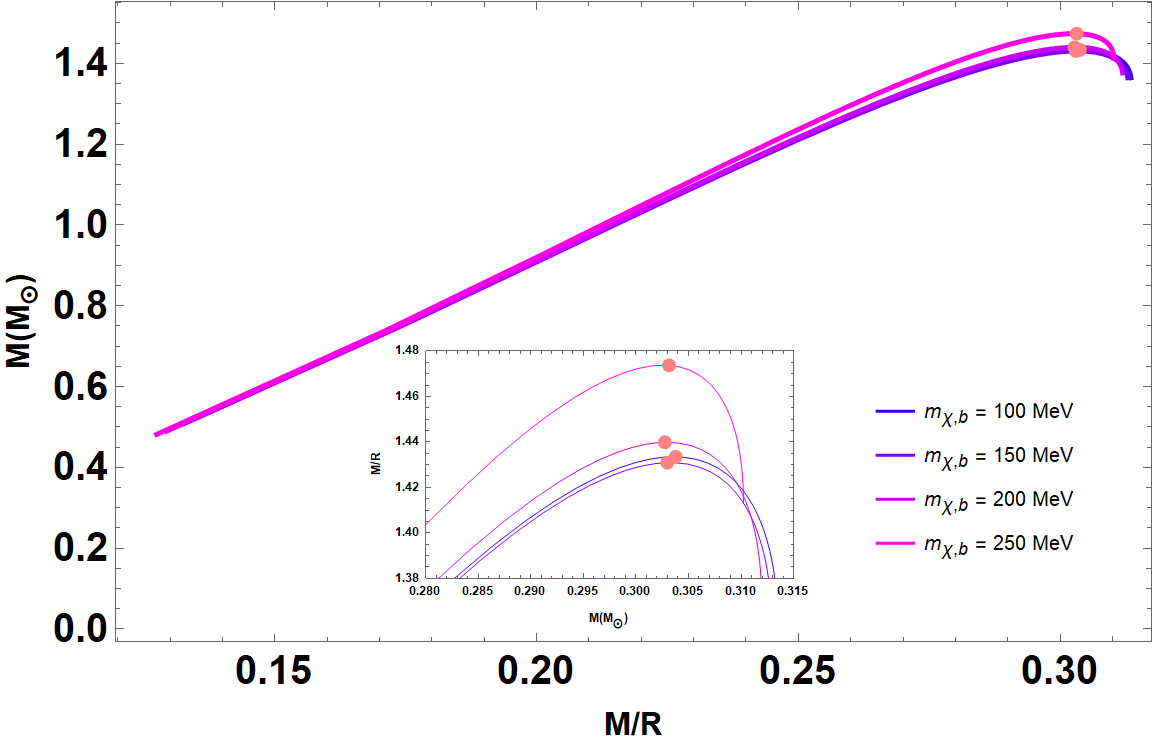}
    \caption{The $(M-R)$ and  $(M-M/R)$ profiles for DM-admixed compact stars with different values of the dimensionless constant $m_{\chi,b} \in [100, 250]$ MeV.  The other parameter sets are $F_x = 0.95$ MeV, $\alpha = 0.5$, and $\rho_s = 1.17 \times 10^{15}$ g/cm $^{3}$, respectively.}
    \label{fig_vary_DM}
\end{figure}

\begin{figure}[h]
    \centering
    \includegraphics[width = 7.1 cm]{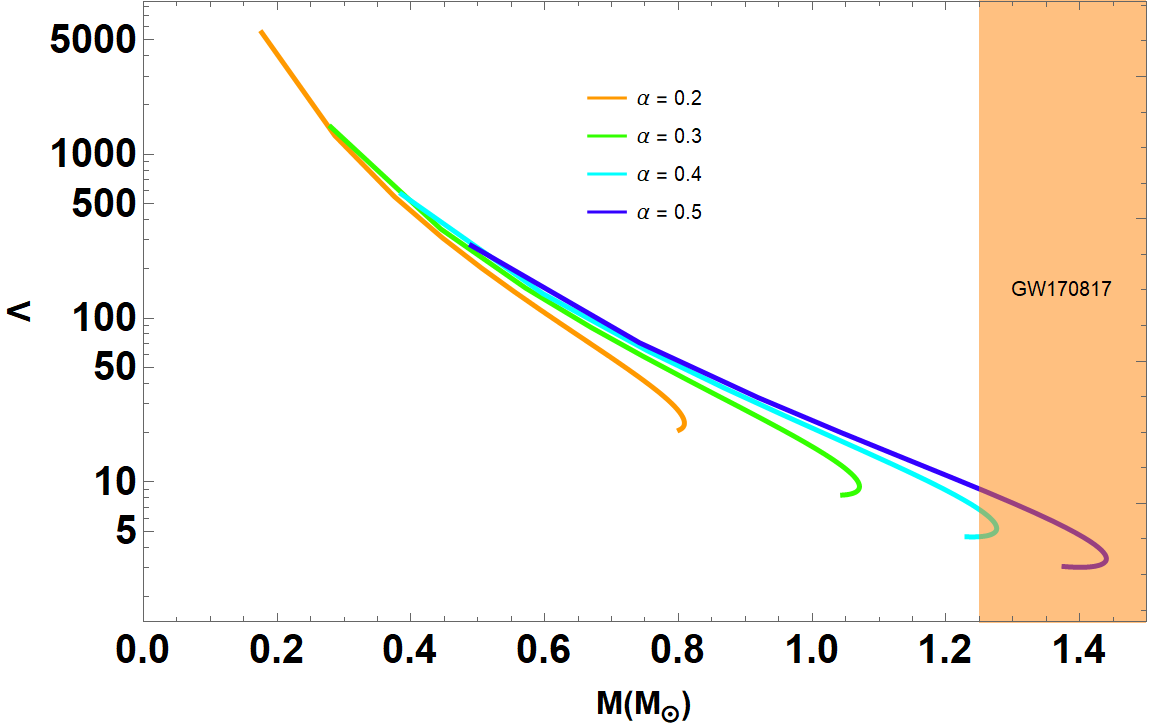}
    \includegraphics[width = 7.1 cm]{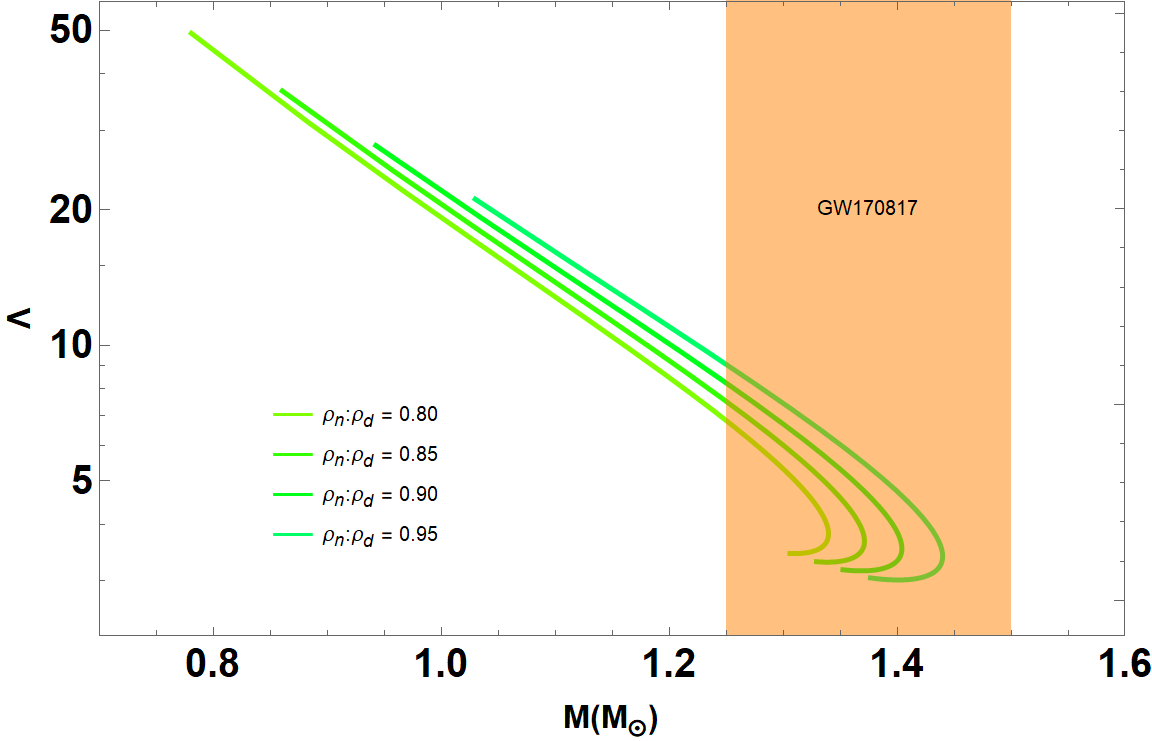}
    \includegraphics[width = 7.1 cm]{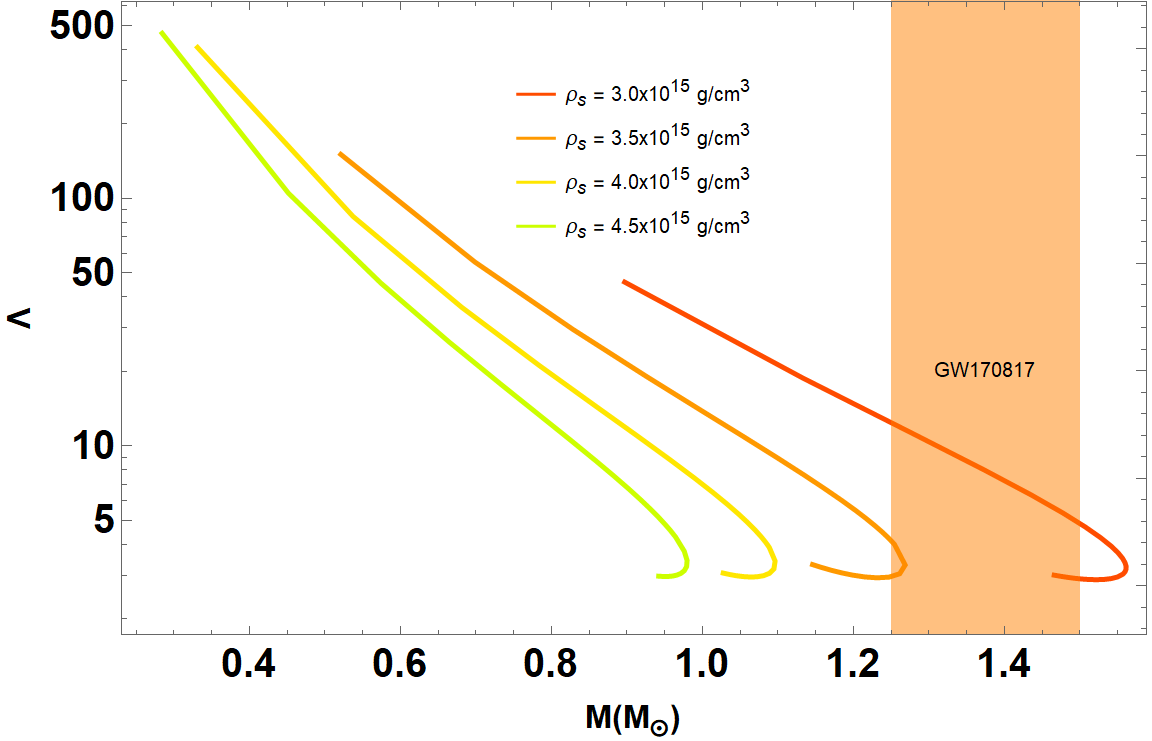}
    \includegraphics[width = 7.1 cm]{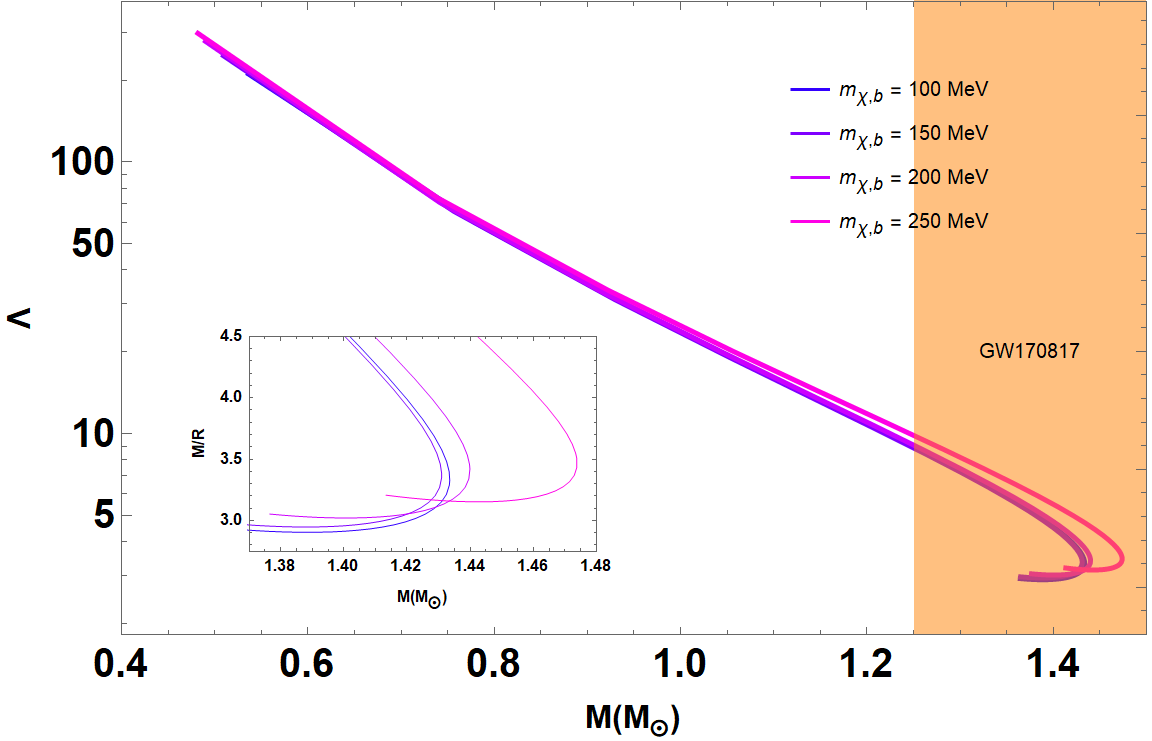}
    \caption{The profiles of tidal deformation and the total mass of the compact stars. We used the same parameter sets as those of Figs. \ref{fig_vary_alpha} to \ref{fig_vary_DM}.}
    \label{fig_lambda}
\end{figure}

\begin{figure}[h]
    \centering
    \includegraphics[width=7.1cm]{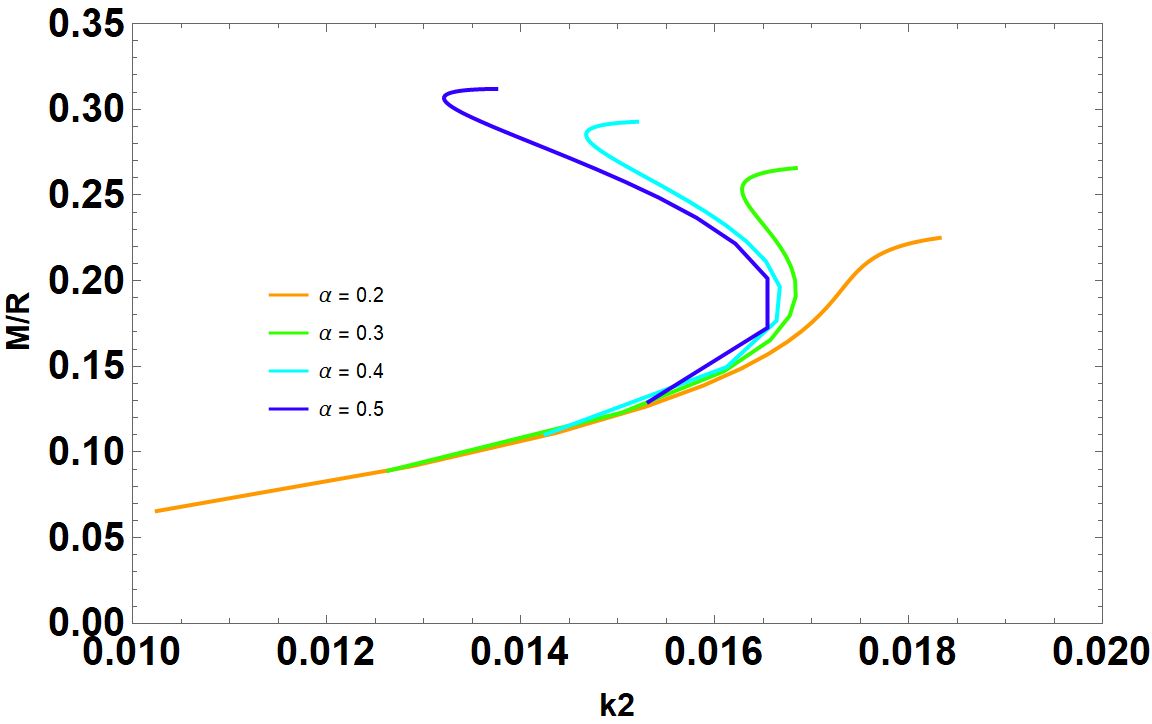}
    \includegraphics[width=7.1cm]{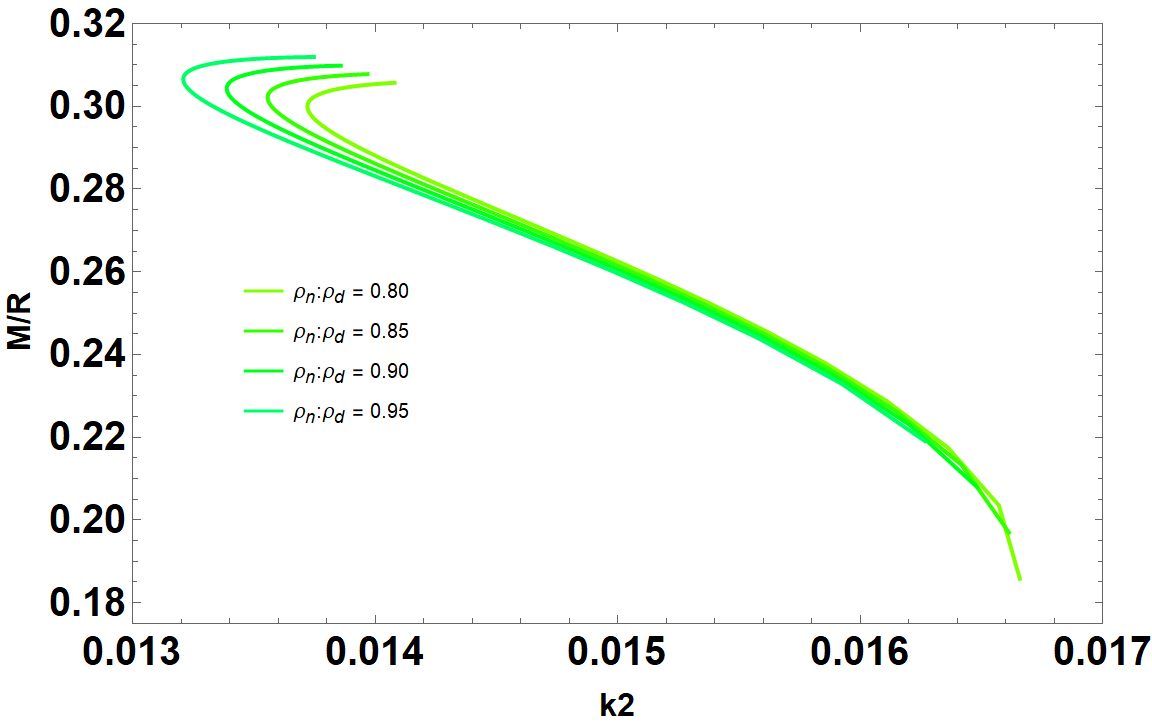}
    \includegraphics[width=7.1cm]{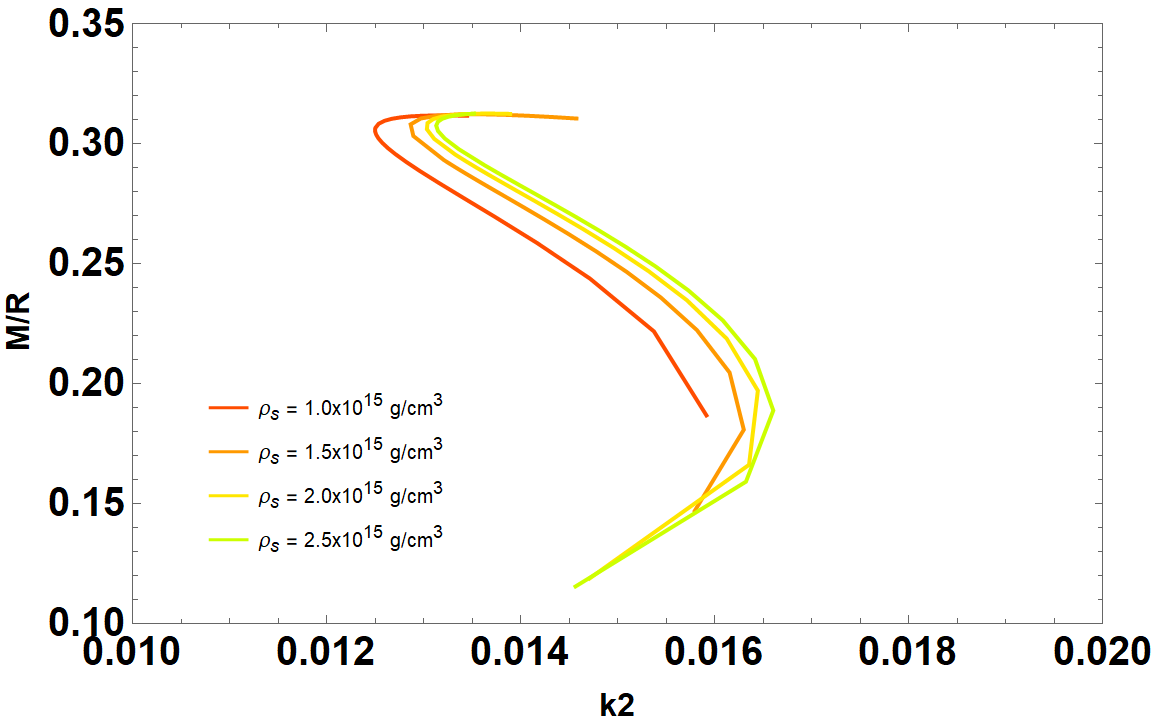}
    \includegraphics[width=7.1cm]{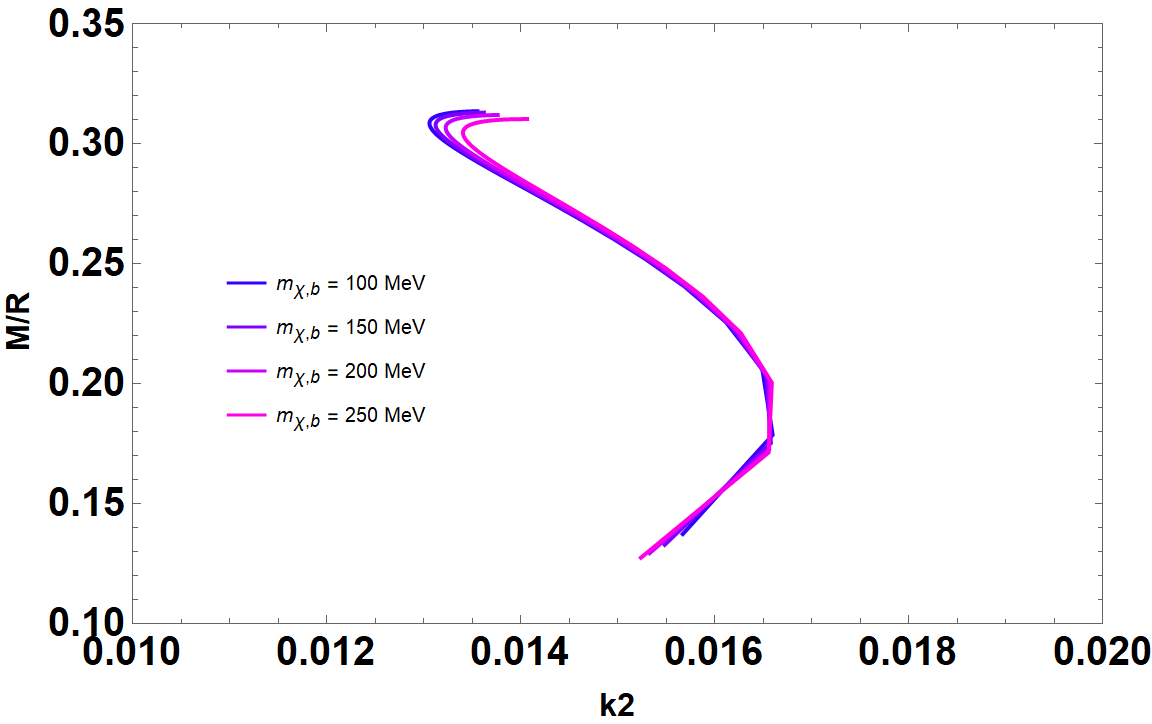}
    \caption{The profiles of $k_2$ and compactness.  We used the same parameter sets as those of Figs. \ref{fig_vary_alpha} to \ref{fig_vary_DM}.}
    \label{fig:enter-label}
\end{figure}

\subsection{Profiles for Variation of $\alpha$}\label{case1}

As a first step, we focus on the impact of the dimensionless constant $\alpha$ on the gravitational mass and radius of CSs in detail. In  Fig.~\ref{fig_vary_alpha}, we illustrate the 
mass-radius $(M-R)$ and mass-compactness $(M-M/R)$ relations
for $\alpha \in [0.2, 0.5]$, while keeping the other parameters fixed as $m_{\chi,b} = 200$ MeV, $F_x= \rho_N/ \rho_D = 0.95$ and $\rho_s = 1.17 \times 10^{15}$ g/cm ${}^{3}$. Observing the $(M-R)$ diagram, we see that $\alpha$ can influence the maximum gravitational mass of CSs, and at $\alpha = 0.5$, this maximum mass is $M_{\rm{max}} = 1.44 M_{\odot}$ with a corresponding radius of $R = 7.05$~km. These results predict that as the value of $\alpha$ increases, the resulting CSs are predicted to be more massive. We explore the constraints on the maximum mass that can be obtained from the NS mergers GW170817 \cite{LIGOScientific:2018cki}. We also plot $(M-M/R)$ relations in the lower panel of Fig.~\ref{fig_vary_alpha}. Here, we observe that the maximum compactness of the star, represented by $M/R$, increases with $\alpha$, and at $\alpha = 0.5$, this value could be $M/R= 0.303$.  A comprehensive overview of the maximum gravitational mass, radius, central density, and compactness is listed in Table \ref{table1}. Interestingly, we find that the values of $M/R$ satisfy the Buchdahl limit i.e., $M/R < 4/9$ \cite{Buchdahl:1959zz}. Overall, we can say that the value of $\alpha$ has a significant impact on the star's overall mass and size while maintaining the stability criterion.


\begin{table}
\caption{\label{table1} Structural properties of compact stars admixed with self-interacting bosonic dark matter  in variations of $\alpha \in [0.2, 0.5]$, while keeping the other parameters fixed as $m_{\chi,b} = 200$ MeV, $F_x= \rho_N/ \rho_D = 0.95$ and $\rho_s = 1.17 \times 10^{15}$ g/cm $^{3}$. The findings are depicted in  Fig. \ref{fig_vary_alpha}.}
    \begin{ruledtabular}
    \begin{tabular}{ccccc}
    $\alpha$  & $M$ [$M_\odot$]  &  $R_{M}$  [\text{km}]  & $\rho_c$ [$\text{MeV/fm}^3$] & $M/R$  \\
    \hline
        0.2 &  0.81 & 5.45 & 4,249 & 0.220 \\
        0.3 &  1.07 & 6.15 & 3,560 & 0.258 \\
        0.4 &  1.28 & 6.66 & 3,147 & 0.284 \\
        0.5 &  1.44 & 7.05 & 2,870 & 0.303 
    \end{tabular}
    \end{ruledtabular}
\end{table}


\begin{table}
\caption{\label{table2} Structural properties of compact stars admixed with self-interacting bosonic dark matter  in variations of $F_x \in [0.8, 0.95]$, while keeping the other parameters fixed as $m_{\chi,b} = 200$ MeV, $\alpha = 0.5$ and $\rho_s = 1.17 \times 10^{15}$ g/cm $^{3}$. The findings are depicted in  Fig. \ref{fig_vary_Ratio}.}
\begin{ruledtabular}
\begin{tabular}{ccccc}
$F_x$  & $M$ [$M_\odot$]  &  $R_{M}$  [\rm{km}]  & $\rho_c$ [MeV/fm$^3$] & $M/R$  \\
\colrule
0.80 &  1.34  &  6.66  &  3,422  &  0.298  \\
0.85 &  1.37  &  6.78  &  3,238  &  0.300   \\
0.90 &  1.40  &  6.91  &  3,054  &  0.301   \\
0.95 &  1.44  &  7.05  &  2,871  &  0.303   
\end{tabular}
\end{ruledtabular}
\end{table}


\begin{table}
\caption{\label{table3} Structural properties of compact stars admixed with self-interacting bosonic dark matter in variations of $\rho_s \in [1.0, 2.5] \times 10^{15}$ g/cm $^{3}$, while keeping the other parameters fixed as $m_{\chi,b} = 200$ MeV, $\alpha = 0.5$ and $F_x= 0.95$. The findings are depicted in  Fig. \ref{fig_vary_rhos}.}
\begin{ruledtabular}
\begin{tabular}{ccccc}
$\rho_s (\times 10^{15})$  & $M$  &  $R_{M}$   & $\rho_c$ & $M/R$  \\
 $[\text{g/cm}^{3}]$ & [$M_\odot$] & [\rm{km}] & [MeV/fm$^3$] & \\
\colrule
1.0 &  1.56  &  7.66  &  2,420  &  0.303  \\
1.5 &  1.27  &  6.20  &  3,709  &  0.303  \\
2.0 &  1.10  &  5.38  &  4,722  &  0.303  \\
2.5 &  0.98  &  4.81  &  5,899  &  0.303   
\end{tabular}
\end{ruledtabular}
\end{table}


\begin{table}
\caption{\label{table4}Structural properties of compact stars admixed with self-interacting bosonic dark matter  in variations of $m_{\chi,b} \in [100, 250]$, while keeping the other parameters fixed as $\alpha = 0.5$, $F_x= 0.95$ and $\rho_s = 1.17 \times 10^{15}$ g/cm $^{3}$. The findings are depicted in  Fig. \ref{fig_vary_DM}.} 
\begin{ruledtabular}
\begin{tabular}{ccccc}
$m_{\chi,b}$   & $M$  &  $R_{M}$    & $\rho_c$  & $M/R$  \\
 $[\text{MeV}]$ & [$M_\odot$] & $[\rm{km}]$ & [MeV/fm$^3$]\\
\colrule
100 &  1.43  &  6.99  &  2,814  &  0.304  \\
150 &  1.43  &  7.00  &  2,814  &  0.303  \\
200 &  1.44  &  7.05  &  2,870  &  0.303  \\
250 &  1.47  &  7.21  &  3,008  &  0.303   
\end{tabular}
\end{ruledtabular}
\end{table}


\subsection{Profiles for variation of $F_x = \frac{\rho_N}{\rho_D}$} \label{case2}

Next, we study the impact of dark matter mass fractions $F_x = \rho_N/\rho_D$ on the $(M-R)$ and  $(M-M/R)$ relations. In Fig. \ref{fig_vary_Ratio}, the $F_x$ is varied between $F_x = 0.8$ and $F_x = 0.95$, while the other parameters $m_{\chi,b} = 200$ MeV, $\alpha = 0.5$ and $\rho_s = 1.17 \times 10^{15}$ g/cm ${}^{3}$ are kept constant. With increasing values of $F_x$, the mass of the star also increases.  At $F_x = 0.95$, the maximum mass is $M_{\rm{max}} = 1.44 M_{\odot}$, with a corresponding radius of $R = 7.05$~km, see Table \ref{table2}. Additionally, the $(M-R)$ profiles are completely consistent with the permitted regions from the analysis of GW170817. To further investigate, we plot $(M-M/R)$ relations in the lower panel of Fig.~\ref{fig_vary_Ratio}. Here, we observed that the maximum 
compactness of the star increases with increasing values of $F_x$, and reaching $M/R= 0.303$ at $F_x = 0.95$. A detailed summary is given in Table \ref{table2} for different values of $F_x$, which reflect that the values of $M/R$ also satisfy the Buchdahl limit i.e., $M/R < 4/9$.

\subsection{Profiles for variation of $\rho_s$} \label{case3}

We have further explored the parameter space of surface density $\rho_s$ concerning the stellar $(M-R)$ and  $(M-M/R)$ relationships. With the fixed values of $m_{\chi,b} = 200$ MeV, $\alpha = 0.5$ and $F_x = 0.95$, we varied $\rho_s$ within the range of $[1.0, 2.5] \times 10^{15}$ g/cm ${}^{3}$. As illustrated in Fig. \ref{fig_vary_rhos}, the maximum mass of the star decreases with an increase in the value of $\rho_s$. As seen in Table \ref{table3}, the maximum gravitational mass could reach  the value of $M_{\rm max} =1.56 M_\odot$ with a corresponding radius $R=7.66$ km at $\rho_s = 1.0 \times 10^{15}$ g/cm ${}^{3}$. Furthermore, we demonstrate that the theoretically attainable maximum mass complies with the constraints of the GW170817 event at the lowest value of the surface energy density. Once more, we have plotted $(M-M/R)$ diagram in the lower panel of Fig. \ref{fig_vary_rhos} depending on the same values. In these cases, we find that the highest value of compactness does not change with the value of $\rho_s$, maintaining a constant ratio of $\frac{M}{R} \sim 0.303$, as shown in Table \ref{table3} and satisfy the Buchdahl limit as well, i.e., $M/R < 4/9$.

\subsection{Profiles for variation of $m_{\chi,b}$} \label{case4}

Finally, we proceed to  study the compliance of DM admixed CSs concerning the variation in the particle
mass of self-interacting bosonic DM. The upper panel of Fig. \ref{fig_vary_DM} is obtained for $\alpha = 0.5$, $F_x = 0.95$ and $\rho_s = 1.17 \times 10^{15}$ g/cm $^{3}$  with different values of $m_{\chi,b} \in [100, 250]$ MeV. It is seen that by increasing $m_{\chi,b}$, both the maximum masses and their corresponding radii also increase. These findings are detailed in Table \ref{table4}, where we see that the maximum achievable mass is $M_{\rm max}$ = $1.47 M_{\odot}$ when $m_{\chi,b}= 250$ MeV, with a corresponding maximum radius of $R_{\rm max}=7.21$ km. We further note that the derived $(M-R)$ relations are in agreement with the GW170817 event, as is shown in Fig. \ref{fig_vary_DM} for each of these models. Finally, we present the effect of $m_{\chi,b}$ on the properties of the $(M-M/R)$ relations. As evident from Fig. \ref{fig_vary_DM} (lower panel) and the data in Table \ref{table4},  we see that the maximum compactness value remains approximately unchanged with respect to variations in $m_{\chi,b}$. The resulting value of $M/R \simeq 0.303$ indicates that $M/R < 4/9$, thereby satisfying the Buchdahl limit for stable configurations.

\section{Results for the tidal deformability}\label{sec5}

The consideration of self-gravitating relativistic objects requires a relativistic theory of tidal deformability. This theory was developed in \cite{flanagan,hinderer,Hussain:2025lqf,damour,Lattimer,poisson}.

We consider a static, spherically symmetric star that is influenced by an external gravitational field, denoted as $\Phi_{ext}$, produced, for instance,  by a companion star in a binary system. The star's reaction to the external field is a deformation, which is predominantly characterized by the development of a quadrupolar moment. $Q_{ij}$ 
\begin{equation}
Q_{ij} = \int d^3x \delta \rho(\vec{x}) \: (x_i x_j - \frac{1}{3} r^2 \delta_{ij}),
\end{equation}
which is proportional to the static external quadrupolar tidal field $E_{ij}$
\begin{equation}
Q_{ij} = - \lambda \: E_{ij},
\end{equation}
\begin{equation}
E_{ij} = \frac{\partial^2 \Phi_{ext}}{\partial x^i \partial x^j},
\end{equation}
and the spatial indices assume three values: $i,j=1,2,3$.

The tidal Love number $k_2$ of a star, a dimensionless coefficient associated with the quadrupole moment, is influenced by the internal structure of perfectly elastic bodies including its mass and EoS. In this formalism, it is directly related to two auxiliary quantities commonly known as `deformabilities', denoted $\lambda$ (dimensionful) and $\Lambda$ (dimensionless), and is given by
\begin{align}
\lambda &\equiv \frac{2}{3} k_2 R^5,
\label{eq:Love1}
\\
\Lambda &\equiv \frac{\lambda}{M^5} = \frac{2 k_2}{3 C^5},
\label{eq:Love2}
\end{align}
where $C=M/R$ is the compactness of the object. Keeping up with traditional conventions, the tidal Love number is expressed in terms of $C$ as follows \cite{flanagan,hinderer,Hussain:2025lqf,damour,Lattimer,poisson}:
\begin{align}
k_2 &= \frac{8C^5}{5} \: \frac{K_{o}}{3  \:K_{o} \: \ln(1-2C) + P_5(C)} ,
\label{elove}
\\
K_{o} &= (1-2C)^2 \: [2 C (y_R-1)-y_R+2] ,
\\
y_R &\equiv y(r=R) ,
\end{align}
where $P_5(C)$ is a fifth-order polynomial given by
\begin{align}
\begin{split}
P_5(C) = \: & 2 C \: \Bigl( 4 C^4 (y_R+1) + 2 C^3 (3 y_R-2) \ + 
\\
&
2 C^2 (13-11 y_R) + 3 C (5 y_R-8) -
\\
&
3 y_R + 6 \Bigl)  ,
\end{split}
\end{align}
%
%
and the function $y(r)$ is the solution of a Riccati differential equation \cite{Lattimer}:
\begin{align}
\begin{split}
r y'(r) + y(r)^2 &+ y(r) e^{\lambda (r)} \Bigl[1 + 4 \pi r^2 ( p(r) - \rho (r) ) \Bigl] 
\\
&+ r^2 Q(r) = 0 ,
\end{split}
\end{align}
supplemented by the initial condition at the center, $r \rightarrow 0$, $y(0)=2$, where the function $Q(r)$, not to be confused with the tensor $Q_{ij}$, is given by
\begin{align}
\begin{split}
  \displaystyle Q(r) = 4 \pi e^{\lambda (r)} \Bigg[ 
  5 \rho (r) 
  &+ 9 p(r) + \frac{\rho (r) + p(r)}{c^2_s(r)} 
  \Bigg] 
  \\
  &- 6 \frac{e^{\lambda (r)}}{r^2} - \Bigl[\nu' (r)\Bigl]^2 .
\end{split}
\end{align}
while $c_s^2 \equiv dp/d\rho = p'(r)/\rho'(r)$ is the speed of sound. 

\smallskip

It is easy to see that since $k_2 \propto (1-2C)^2$, tidal Love numbers of black holes vanish due to the fact that the factor of compactness of black holes $C=1/2$. This is an intriguing result of classical GR saying that tidal Love numbers of black holes, as opposed to other types of compact objects, are precisely zero. Therefore, a measurement of a non-vanishing $k_2$ will be a smoking-gun deviation from the standard black hole of GR.

\smallskip

In Fig. \ref{fig_lambda} we have displayed the dimensionless deformability $\Lambda$ versus stellar mass (in solar masses). It is a rapidly decreasing function that takes large values for light stars and small values for massive stars. We have varied each considered case and observed that $\Lambda$ increases with $\alpha, F_x, m_{\chi,b}$ and decreases with $\rho_s$.  We find that in all cases the constraint for the canonical star (i.e. object of stellar mass $M=1.4 M_{\odot}$) coming from the GW170817 event $\Lambda_{1.4} < 800$ is satisfied \cite{GW170817}. Our results indicate that the impact of dark matter on stellar properties-via the variation of $m_{\chi,b}, F_x$-is as follows: Both a higher $F_x$ and a more massive DM particle lead to a larger highest stellar mass (which is equivalent to a stiffer EoS), as shown in Tables and Figures, and consequently this implies a larger deformability.

Moreover, we have shown the gravito-electric tidal Love numbers $k_2$ versus factor of compactness $C=M/R$ in Fig. \ref{fig:enter-label}. The tidal Love numbers first increase with $M/R$, they reach a maximum value, and after that they decrease. This is to be expected, since as $k_2 \propto (1-2 C)^2$, it tends to zero as $C \rightarrow 1/2$. Here, too, we have varied $\alpha, \rho_s, m_{\chi,b}, F_x$ and the results are displayed in the four panels. This time we observe that the tidal Love numbers increase with $\rho_s,m_{\chi,b}$ and decrease with $\alpha,F_x$.

\section{Stability Analysis of compact stars}\label{sec6}

This section is dedicated to assessing the stability of our proposed model. To achieve this evaluation, we utilize the static stability criterion, the adiabatic index, and the sound speed. Each condition related to stability is systematically examined and illustrated graphically.

\subsection{Static Stability Criterion}

\begin{figure}[h]
    \centering
    \includegraphics[width = 8.3 cm]{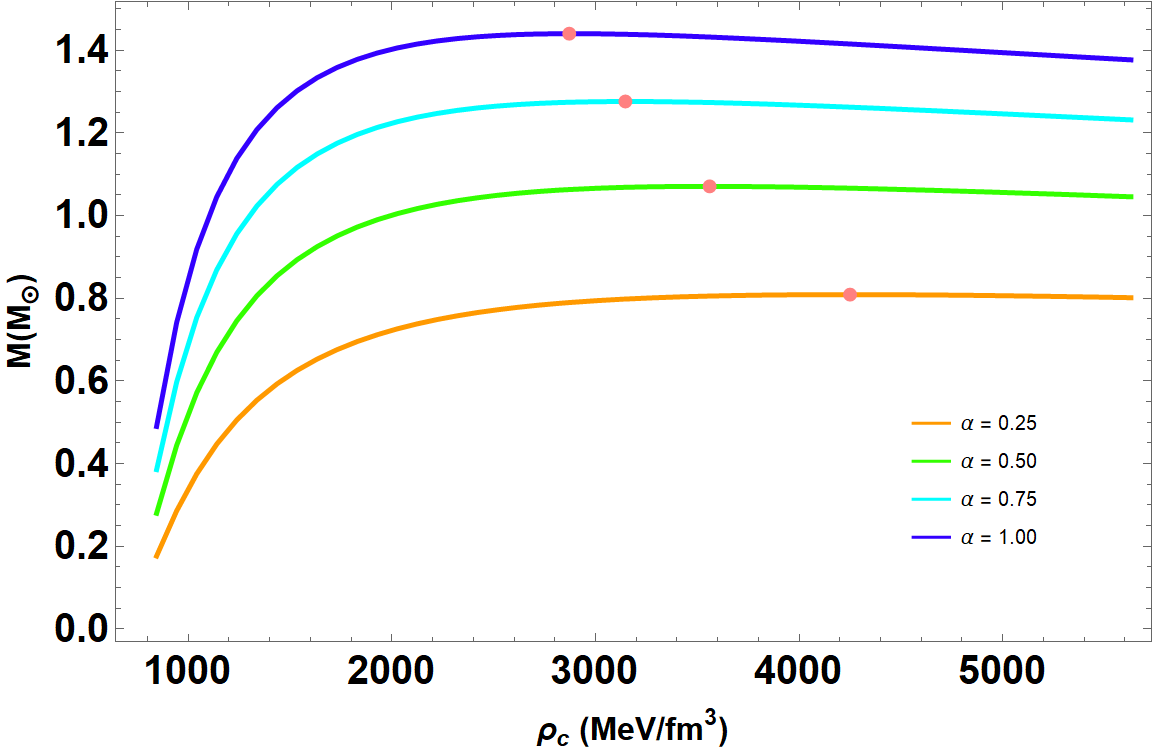}
    \includegraphics[width = 8.3 cm]{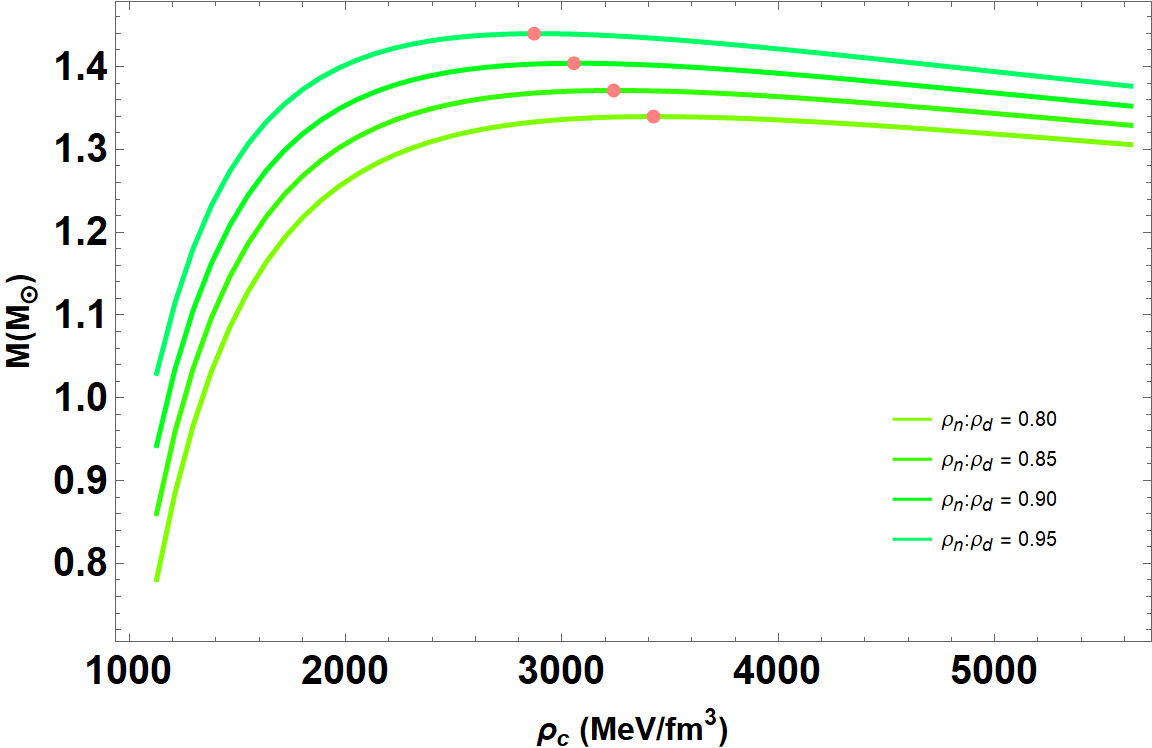}
    \includegraphics[width = 8.3 cm]{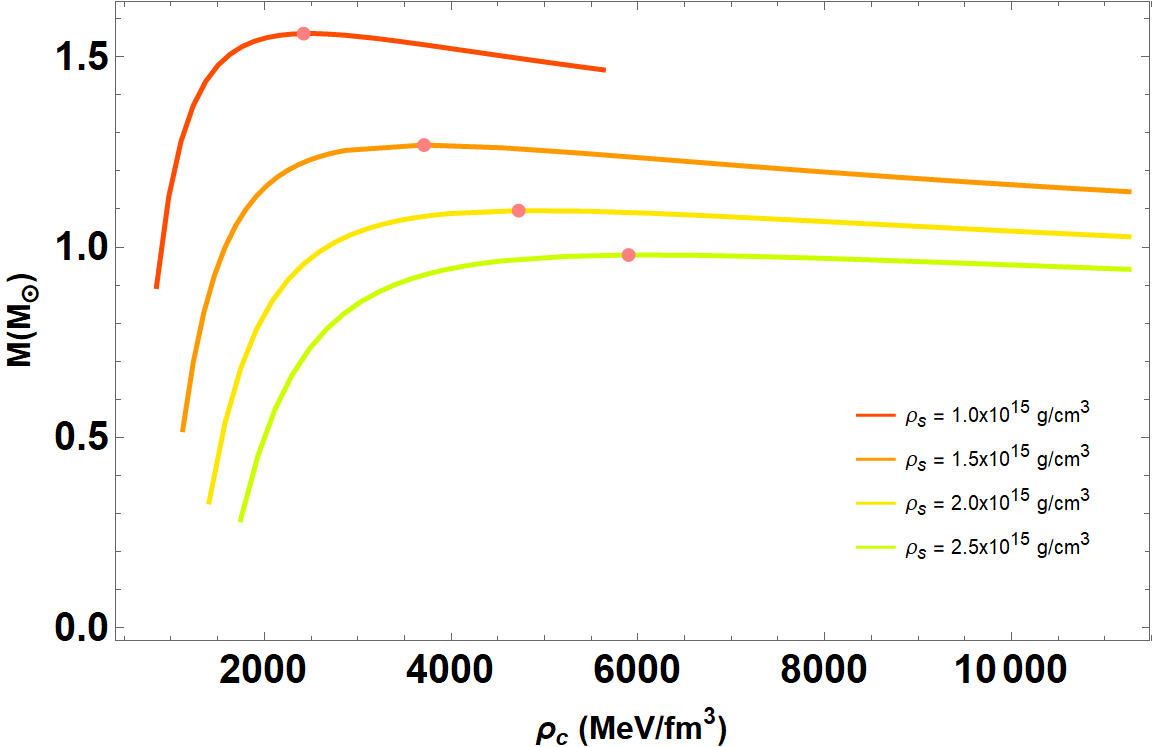}
    \caption{This profile is for the maximum mass $M$ versus the central energy density $\rho_c$.  We used the same parameter sets as those of Figs. \ref{fig_vary_alpha} to \ref{fig_vary_DM}.}
    \label{sss}
\end{figure}

In this investigation, we analyze the stability of equilibrium configurations emphasizing the \textit{static stability criterion} outlined in Refs. \cite{harrison,ZN}.
The outcomes are illustrated in the $M-\rho_c$ plane, where $M$ indicates the gravitational mass and $\rho_c$ refers to the central energy density. The condition is expressed as:
\begin{eqnarray}
\frac{dM}{d\rho_c} < 0 &~ \rightarrow \text{unstable configuration}, \\
\frac{dM}{d\rho_c} > 0 &~ \rightarrow \text{stable configuration}.
\label{criterion_M_rho_c}
\end{eqnarray}
It is important to highlight that while this condition is necessary for stability, it is not sufficient on its own. 
The relationships between mass and central density are depicted in Fig.~\ref{sss} for each scenario examined. 
In those figures, the pink points $(M_{\text{max}}, R_{M_{\text{max}}})$ serve as a boundary point that separates the region of stability from that of instability, and this is indicated by $dM/d\rho_c = 0$, while $dM/d\rho_c > 0$ signifies stability.


\begin{figure}[h]
    \centering
    \includegraphics[width = 7.0 cm]{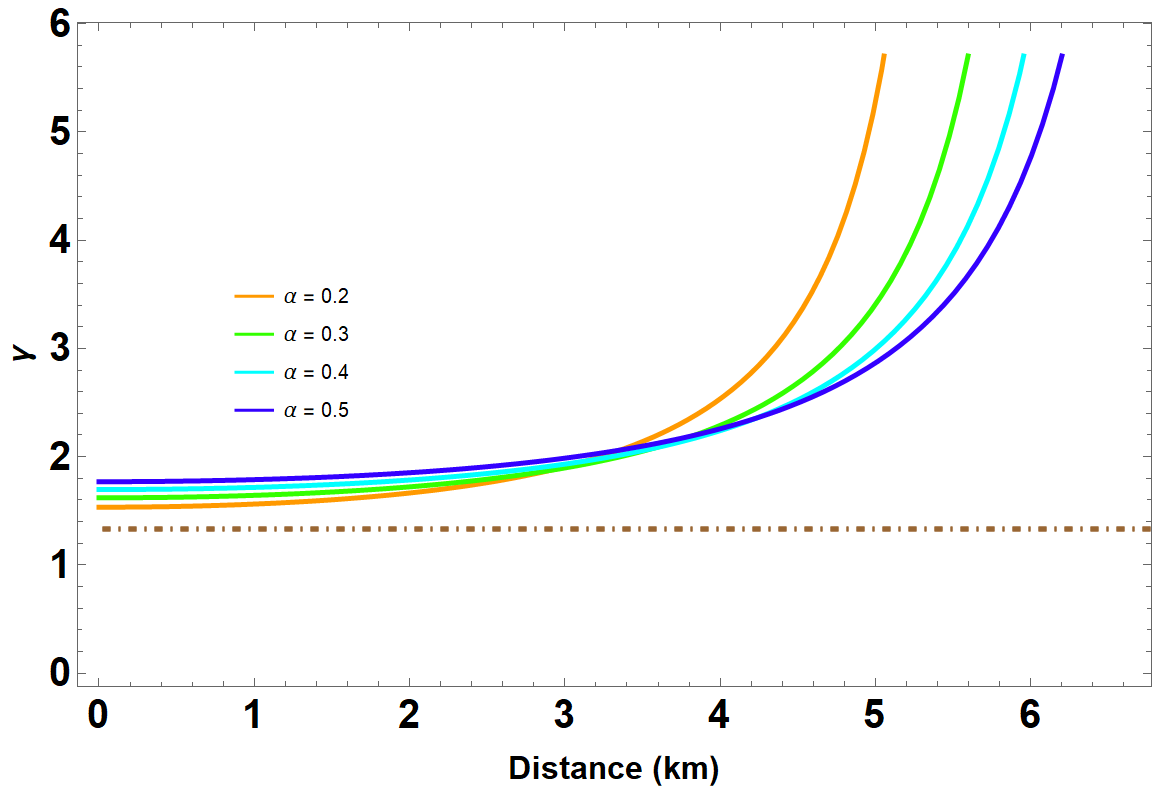}
    \includegraphics[width = 7.0 cm]{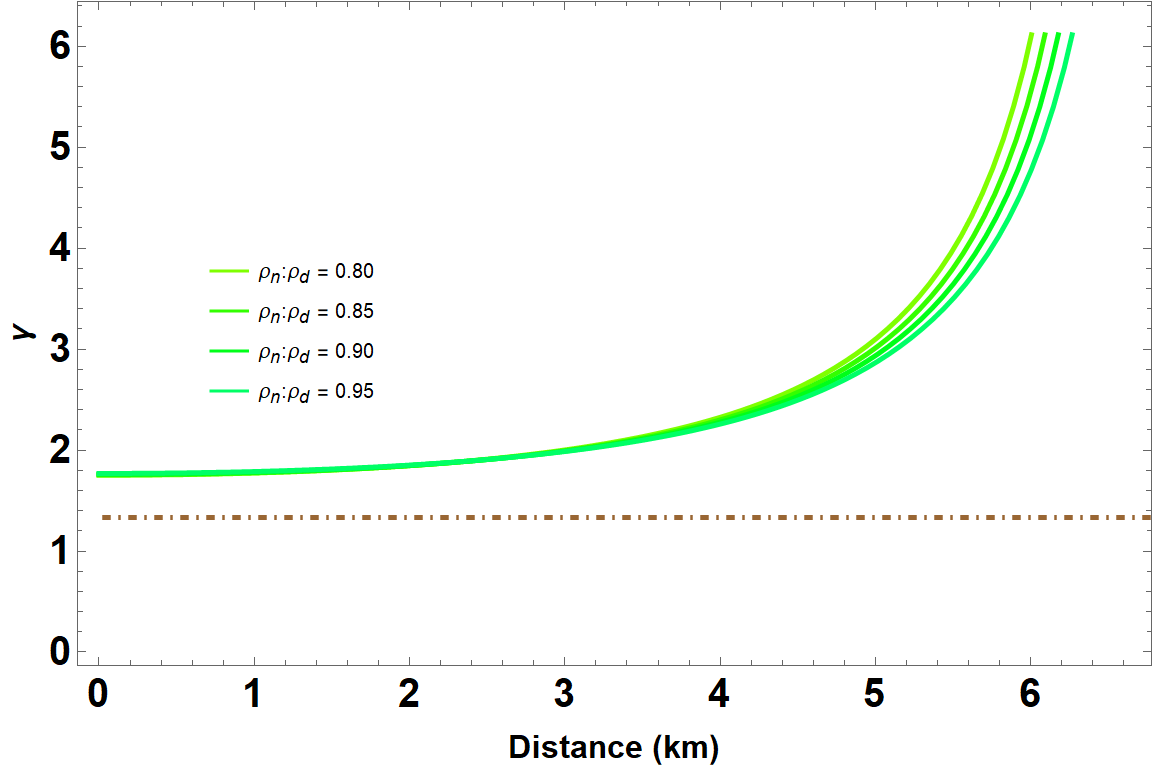}
    \includegraphics[width = 7.0 cm]{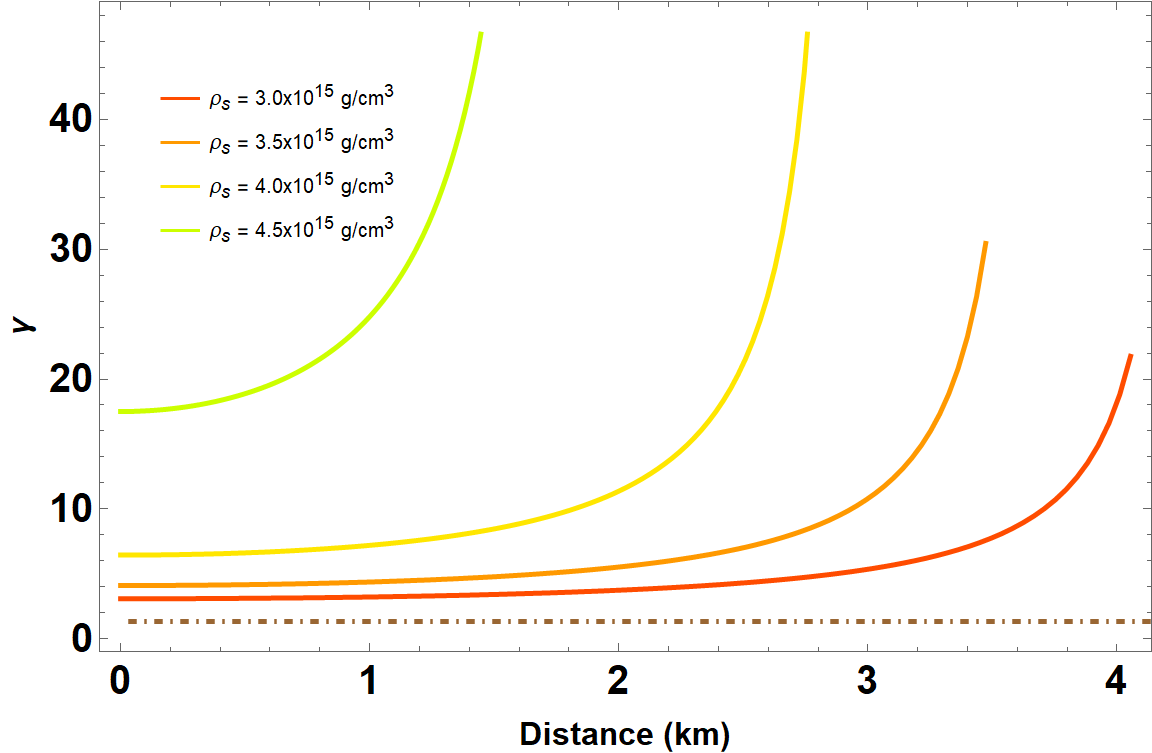}
    \includegraphics[width = 7.0 cm]{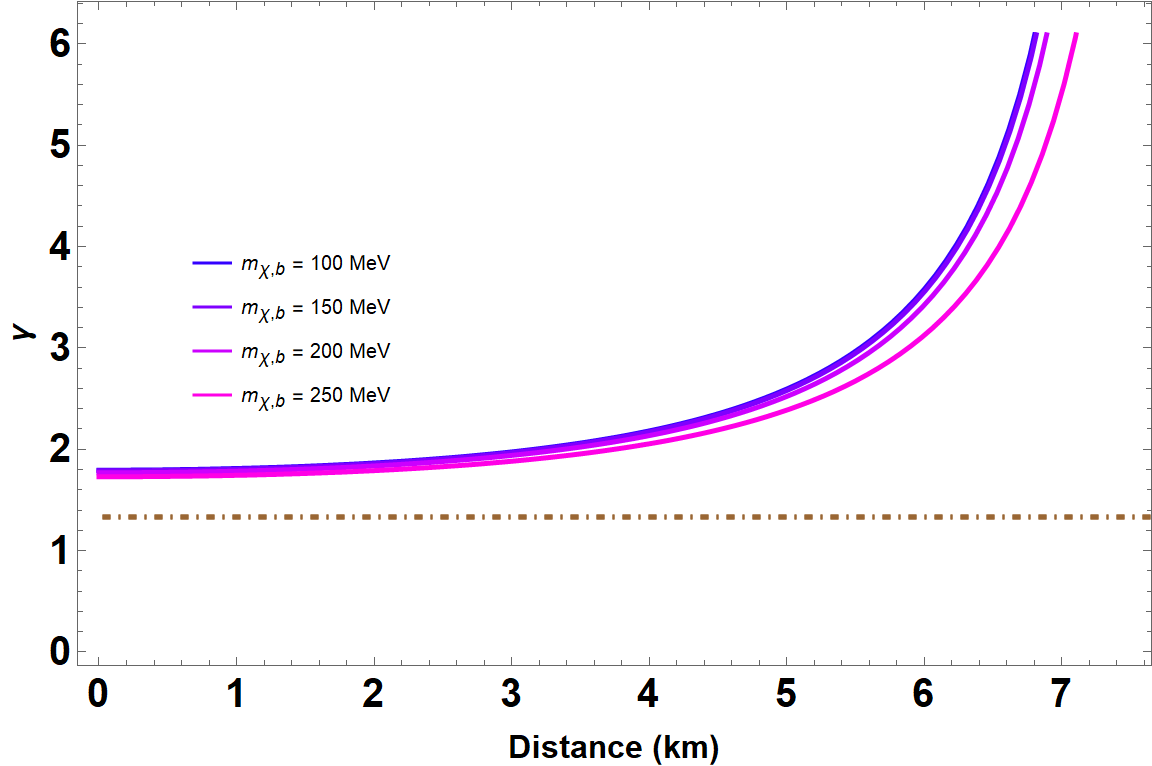}
    \caption{ The adiabatic index ($\gamma$) has been plotted as a function of radial coordinate $r$.  We used the same parameter sets as those of Figs. \ref{fig_vary_alpha} to \ref{fig_vary_DM}.}
    \label{fig_stable_vary_alpha}
\end{figure}


\begin{figure}[h]
    \centering
    \includegraphics[width = 7.5 cm]{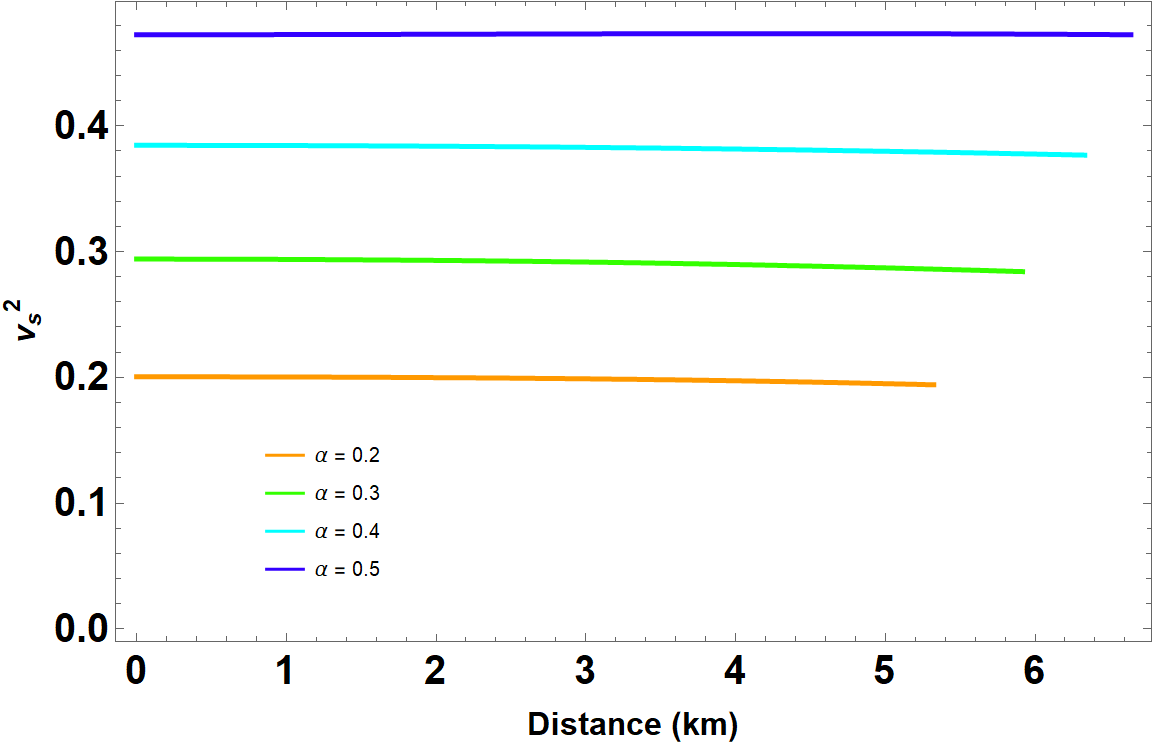}
    \includegraphics[width = 7.5 cm]{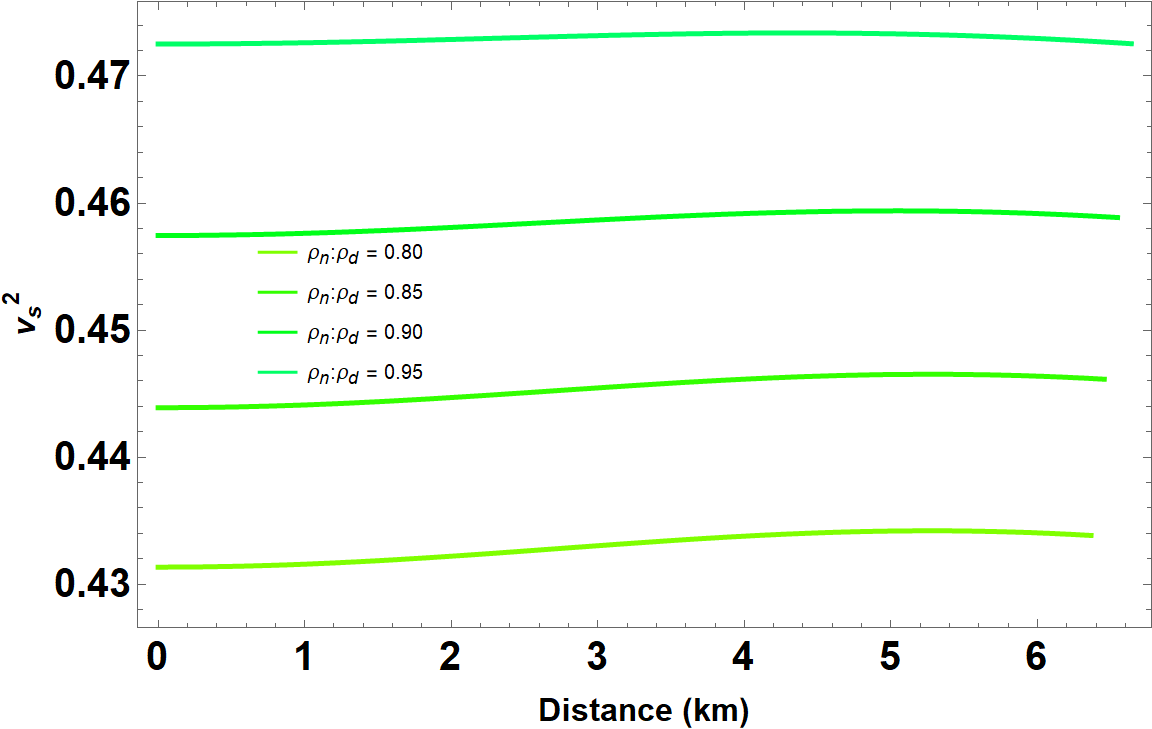}
    \includegraphics[width = 7.5 cm]{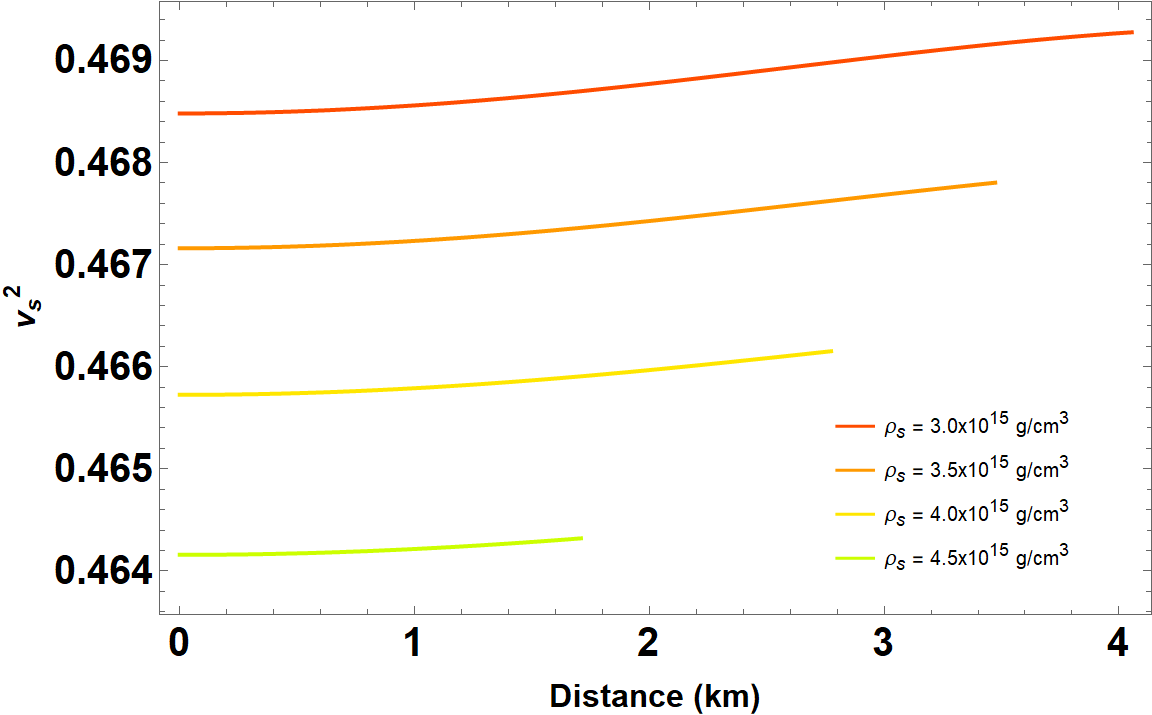}
    \includegraphics[width = 7.5 cm]{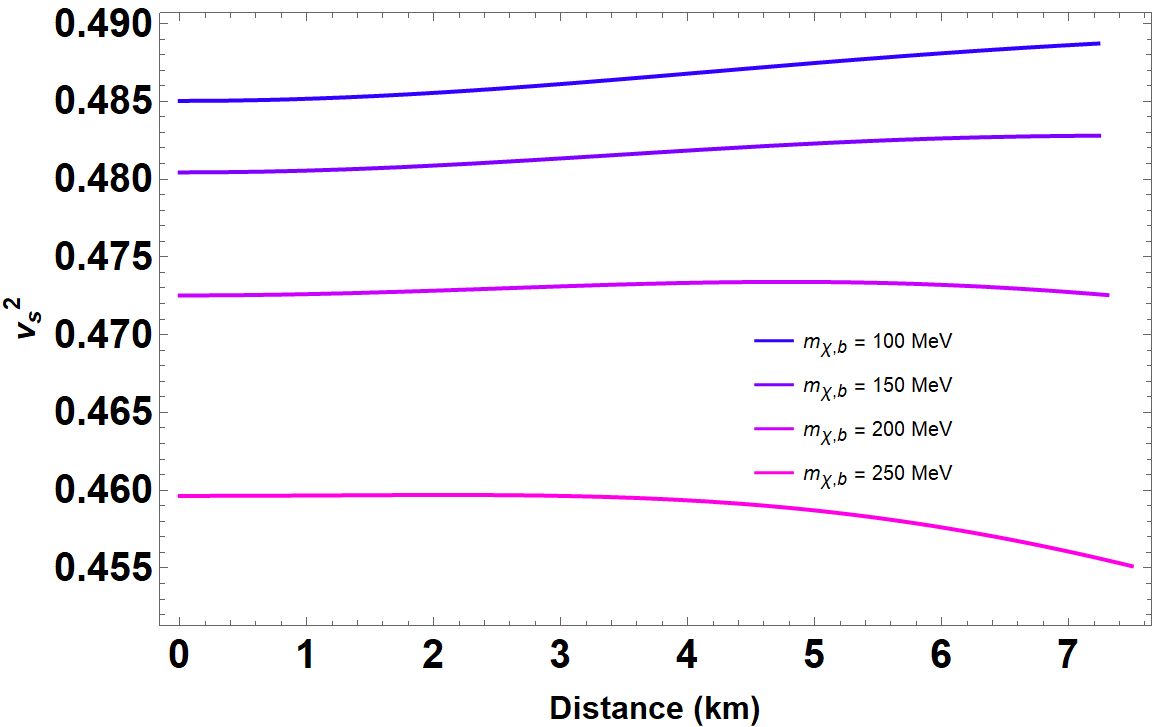}
    \caption{The squared speed of sound for QSs calculated under the same parameter sets as of Figs. \ref{fig_vary_alpha} to \ref{fig_vary_DM}.}
    \label{fig_stable_vary_Ratio}
\end{figure}


\subsection{Adiabatic Indices}

We additionally compute the adiabatic index, $\gamma$, to perform the dynamical stability of the astrophysical models. This notion, originally put forth by Chandrasekhar \cite{Chandrasekhar}, examines the stability of an equilibrium configuration. The radial adiabatic index $\gamma$ is defined as:
\begin{eqnarray}\label{adi}
    \gamma \equiv \left(1 + \frac{\rho}{p}\right) {v_s^2} ,
\end{eqnarray}
where ${v_s^2}=\frac{dp}{d\rho}$ is the sound speed.The stability of a static configuration is determined by the critical value associated with the adiabatic index $\gamma$. This critical value is expressed as $\langle \gamma \rangle > \gamma_{cr}$, where $\langle \gamma \rangle$ denotes the mean value of the relativistic adiabatic index \cite{Moustakidis:2016ndw} (see also Refs. \cite{Malik:2025ntj,Hussain:2025gvx} for recent reviews). In the pure Newtonian gravitational theory,  the value of $\gamma_{cr}$ consistently remains at $4/3$. In contrast, when one incorporates general relativistic effects, the critical value becomes $(4/3) + (19 M)/(21 R)$. By employing Eq.~(\ref{adi}), we depict in Fig.~\ref{fig_stable_vary_alpha} how $\gamma$ varies with the radius for different selected values of $(\alpha, {\rho_N}/{\rho_D}, \rho_s ~\text{and} ~m_{\chi,b})$. The analysis reveals that all the computed cases satisfy the criterion. To this end, we can say that the dark matter-admixed compact stars is stable when subjected to adiabatic conditions.

\subsection{Sound Speed and Causality}

To assess the stability of compact stars, we calculated the speed of sound $v_{s}^2 = \frac{dP}{d\rho}$, which constitutes a fundamental requirement for stability analysis, and it should be less than the speed of light, i.e., $v_{s}^2 < 1$. From Fig. \ref{fig_stable_vary_Ratio} it is evident that all diagrams satisfy this requirement for the variation of $(\alpha, {\rho_N}/{\rho_D}, \rho_s \sim \text{and} \sim m_{\chi,b})$.

\section{Concluding Remarks}\label{sec7}

In this paper, we have comprehensively studied compact stars that are admixed with self-interacting bosonic dark matter. Our findings demonstrate that the inclusion of dark matter in CS significantly affects their structural properties and overall stability. Moreover, we studied the tidal deformability of CSs admixed with DM and compared our results with the tidal deformability data obtained from the GW170817 event.

Within this work our study explores the effects of varying four sets of parameters: (i) $\alpha$ (dimensionless constant related to sound speed), 
(ii) $\rho_s$ (surface density), (iii) $m_{\chi,b}$ (dark matter particle mass) and (iv) $F_x$ (dark matter mass fraction), respectively, and
generate $(M-R)$ and $(M-M/R)$ relations across all possible combinations. Our study found that both the DM particle mass and DM fraction could affect the mass, radius, and overall structural configurations of CSs. With the increase of $m_{\chi,b}$ and $F_x$, the maximum allowed value could reach $1.44 M_{\odot}$, which follows the observational constraints we impose.  Further, our attention has been focused on the variation of surface density $\rho_s$, and the $(M-R)$ relations depending on it. As expected, we find that the maximum mass of the star decreases with
an increase in the value of $\rho_s$. Additionally, we compute the maximum compactness for each scenario discussed in Sec \ref{sec4}. According to our findings, we conjecture
that maximum compactness lies in the range of $0.220 \leq M/R\leq 0.304$, and thus satisfies the bound on the compactness i.e., $C = M/R \leq 4/9  \equiv 0.444$
for physically admissible self-gravitating objects in GR.

Moreover, our investigation also examined the tidal deformation properties of the CSs.  The key property is the dimensionless deformability $\Lambda$ rapidly declining function, which assumes high values for lighter stars and lower values for heavier stars. In all the cases we examined, the stability of the configurations has been analyzed by performing the static stability criterion, the adiabatic index, and the sound velocity.  Our results clearly indicate that DM admixed CSs meet the stability requirement $dM/d\rho_c > 0$ and have adiabatic indices exceeding the critical value of $4/3$, thereby indicating stable configurations.  

The present study offers a theoretical framework for experimental investigations of DM, potentially extending to explore DM density profiles that dynamically evolve in response to NS/QS environments, taking into account both core and crustal effects. Additionally, this framework could help in depicting the characteristics of DM or in further constraining its nature.

\section*{acknowledgments}

T. T. was supported by Walailak University under the New Researcher Development scheme (Contract Number WU67268).  He also acknowledges COST action CA21106 and CA22113.

\end{document}